\documentclass{aa}
\usepackage{graphicx}
\usepackage{txfonts}
\newcommand{\etal}{et\,al.\ }
\newcommand{\logg}{\mbox{$\log g$}}
\newcommand{\Teff}{\mbox{$T_\mathrm{eff}$}}
\newcommand{\hh}{\object{H\,1504$+$65}}

\newcommand{\lppr}{\stackrel{<}{\scriptstyle \sim}}
\newcommand{\lappr}{\raisebox{-0.4ex}{$\lppr $}}
\newcommand{\hse}{\hbox{}\hspace{1.1mm}\hbox{}}
\newcommand{\hsw}{\hbox{}\hspace{1.3mm}\hbox{}}
\newcommand{\hsed}{\hbox{}\hspace{0.8mm}\hbox{}}
\newcommand{\hses}{\hbox{}\hspace{0.4mm}\hbox{}}
\newcommand{\hsx}{\hbox{}\hspace{1.5mm}\hbox{}}
\begin{document}
   \title
   {Chandra and FUSE spectroscopy of the hot bare stellar core \hh
    \thanks
{Based on observations made with the NASA-CNES-CSA Far Ultraviolet 
Spectroscopic Explorer. FUSE is operated for NASA by the Johns Hopkins
University under NASA contract NAS5-32985.
}
   }
 
   \author{K. Werner$^1$, T. Rauch$^{1,2}$, M.A. Barstow$^3$ \and J.W. Kruk$^4$}
   \offprints{K\@. Werner}
   \mail{werner@astro.uni-tuebingen.de}
 
   \institute
    {
     Institut f\"ur Astronomie und Astrophysik, Universit\"at T\"ubingen, Sand 1, D-72076 T\"ubingen, Germany
\and
Dr.-Remeis-Sternwarte, Universit\"at Erlangen-N\"urnberg, Sternwartstra\ss e 7, D-96049 Bamberg, Germany
\and
Department of Physics and Astronomy, University of Leicester, University
    Road,  Leicester, LE1 7RH, UK
\and
Department of Physics and Astronomy, Johns Hopkins University,
Baltimore, MD 21218, U.S.A.
}
    \date{Received 29 January 2004 / Accepted 6 April 2004}
   \authorrunning{K. Werner et al.}
   \titlerunning{Chandra and FUSE spectroscopy of the hot bare stellar core \hh}
   \abstract{\hh\ is an extremely hot hydrogen-deficient white dwarf with an effective
temperature close to 200\,000\,K. We present new FUV and soft X-ray spectra
obtained with FUSE and Chandra, which confirm that \hh\ has an atmosphere
primarily composed of carbon and oxygen. The Chandra LETG spectrum (60\,--\,160\AA) shows a wealth of
photospheric absorption lines from highly ionized oxygen, neon, and -- for the first time
identified in this star -- magnesium and suggests
relatively high Ne and Mg abundances. This corroborates an earlier suggestion
that \hh\ represents a naked C/O stellar core or even the C/O envelope of an
O-Ne-Mg white dwarf.
        }
   \maketitle

\begin{table}
\caption{Summary of the model atoms used in the model atmosphere and line formation calculations.
The numbers in brackets give the individual line numbers summed into superlines
for the heavy metal ions.
\label{levels_tab}}
\small
\begin{tabular}{l l r r r}
      \hline
      \hline
      \noalign{\smallskip}
element & ion & NLTE levels & lines&\\
  \noalign{\smallskip}
 \hline
      \noalign{\smallskip}
      He & \scriptsize I          & 1  & 0  \\
         & \mbox{\scriptsize II}  & 10 & 36 \\ 
         & \mbox{\scriptsize III} & 1  & --  \\ 
      \noalign{\smallskip}
      C  & \mbox{\scriptsize III} & 6  & 4  \\
         & \mbox{\scriptsize IV}  & 36 & 98 \\
         & \mbox{\scriptsize V}   & 1  & 0  \\
      \noalign{\smallskip}
      O  & \mbox{\scriptsize IV}  & 1  & 0  \\
         & \mbox{\scriptsize V}   & 20 & 24 \\
         & \mbox{\scriptsize VI}  & 52 & 231\\ 
         & \mbox{\scriptsize VII} & 1  & 0  \\
      \noalign{\smallskip}
      Ne & \mbox{\scriptsize IV}  & 1  & 0  \\
         & \mbox{\scriptsize V}   & 5  & 0  \\
         & \mbox{\scriptsize VI}  & 23 & 56 \\
         & \mbox{\scriptsize VII} & 54 & 250\\
         & \mbox{\scriptsize VIII}& 5  & 6  \\
         & \mbox{\scriptsize IX}  & 1  & 0  \\
      \noalign{\smallskip}
      Mg & \mbox{\scriptsize IV}  & 1  & 0  \\
         & \mbox{\scriptsize V}   & 10 & 7  \\
         & \mbox{\scriptsize VI}  & 48 & 161\\
         & \mbox{\scriptsize VII} & 39 & 86 \\
         & \mbox{\scriptsize VIII}& 23 & 56 \\
         & \mbox{\scriptsize IX}  & 1  & 0  \\
      \noalign{\smallskip}
      Al & \mbox{\scriptsize IV}  & 1  & 0  \\
         & \mbox{\scriptsize V}   & 7  & 4  \\
         & \mbox{\scriptsize VI}  & 20 & 19 \\
         & \mbox{\scriptsize VII} & 41 & 31 \\
         & \mbox{\scriptsize VIII}& 35 & 16 \\
         & \mbox{\scriptsize IX}  & 1  & 0  \\
       \noalign{\smallskip}
      Na & \mbox{\scriptsize IV}  & 1  & 0  \\
         & \mbox{\scriptsize V}   & 41 & 142\\
         & \mbox{\scriptsize VI}  & 43 & 130\\
         & \mbox{\scriptsize VII} & 43 & 164\\
         & \mbox{\scriptsize VIII}& 11 & 13 \\
         & \mbox{\scriptsize IX}  & 1  & 0  \\
      \noalign{\smallskip}
      Fe & \mbox{\scriptsize VI}  & 7 & 25 & (340\,132)\\
         & \mbox{\scriptsize VII} & 7 & 24 & (86\,504)\\
         & \mbox{\scriptsize VIII}& 7 & 27 & (8\,724)\\
         & \mbox{\scriptsize IX}  & 7 & 25 & (36\,843)\\
         & \mbox{\scriptsize X}   & 7 & 28 & (45\,229)\\
         & \mbox{\scriptsize XI}  & 1 &  0 & \\
           \noalign{\smallskip}
      \noalign{\smallskip}
      Ni & \mbox{\scriptsize VI}  & 7 & 27 & (1\,110\,584)\\
         & \mbox{\scriptsize VII} & 7 & 18 & (688\,355)\\
         & \mbox{\scriptsize VIII}& 7 & 27 & (553\,549)\\
         & \mbox{\scriptsize IX}  & 7 & 24 & (79\,227)\\
         & \mbox{\scriptsize X}   & 1 &  0 & \\
       \noalign{\smallskip}
 generic & \mbox{\scriptsize VI}  & 7 & 28 & (1\,755\,957)\\
 (Ca-Ni) & \mbox{\scriptsize VII} & 7 & 28 & (1\,244\,159)\\
         & \mbox{\scriptsize VIII}& 7 & 28 & (819\,983)\\
         & \mbox{\scriptsize IX}  & 8 & 36 & (844\,432)\\
         & \mbox{\scriptsize X}   & 1 &  0 & \\
           \noalign{\smallskip}
\hline
\normalsize
     \end{tabular}
\normalsize
\end{table}

\section{Introduction} 

\hh\ is a faint blue star that has been identified as the counterpart of a
bright soft X-ray source (Nousek \etal 1986)  discovered by an early X-ray
survey (Nugent \etal 1983). Spectroscopically, the star is a member of the
PG1159 class which comprises hot hydrogen-deficient (pre-) white dwarf stars
($T_{\rm eff}$=75\,000\,K--180\,000\,K, $\log g$=5.5--8 [cgs]; Werner 2001). The
PG1159 stars are probably the outcome of a late helium-shell flash, a phenomenon
that drives the currently observed fast evolutionary rates of three well-known
objects (FG~Sge, Sakurai's object, V605 Aql). A late helium-shell flash may
occur in a post-AGB star or a white dwarf.
Flash induced envelope mixing generates a H-deficient surface layer. The
surface chemistry then essentially reflects that of the region between the H-and
He-burning shells of the precursor AGB star. 
The He-shell flash transforms the star back to an AGB
star and the subsequent, second post-AGB evolution explains the existence of
Wolf-Rayet central stars of planetary nebulae and their successors, the PG1159
stars.

Within the PG1159
group \hh\ is an extraordinary object, as it has been shown that it is not only
hydrogen-deficient but also helium-deficient.  From optical spectra it was
concluded that the atmosphere is primarily composed of carbon and oxygen, by
equal amounts (Werner 1991, {\small W91}). Strong neon lines were detected in soft X-ray spectra
taken with the EUVE satellite and in an optical-UV Keck spectrum and an
abundance of Ne=2--5\% (all abundances in this paper are given as mass
fractions) was derived (Werner \& Wolff 1999, {\small WW99}). The origin of this exotic surface
chemistry is unclear.

Here we present new results of
observations performed with FUSE (Far Ultraviolet Spectroscopic Explorer) and
the Chandra X-ray Observatory, whose spectroscopic resolution is
more than an order of magnitude better than that of previous FUV and EUV
missions (HUT and EUVE). We will show that the Chandra spectrum contains a
wealth of photospheric absorption lines from highly ionized metals. Ionization
balances of O, Ne, and Mg lines provide new constraints on the effective temperature and,
for the first time, allow an estimation of the Mg abundance. We also analyze the
spectra with respect to other metals, namely Al, Na, and the iron group.

In the following sections we will first describe the model atmosphere
calculations. Then we will present in detail the compilation of atomic data and
design of the model atoms. After a coarse characterization of the FUSE and
Chandra spectra we turn to the detailed comparison between synthetic and
observed spectra. Finally we discuss the results and their implications on the
evolutionary history of \hh. 

\begin{figure}[tbp]
  \resizebox{\hsize}{!}{\includegraphics{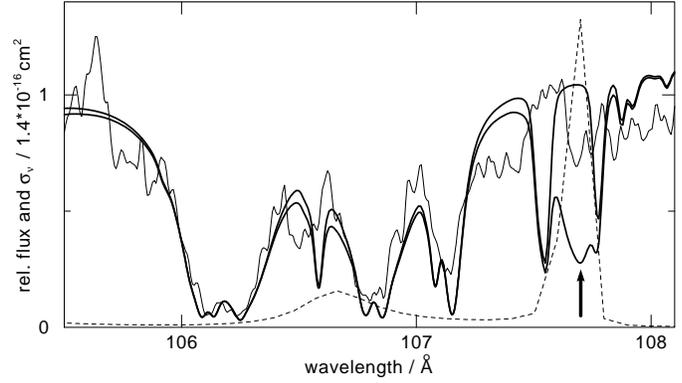}}
  \caption[]{
Chandra observation (thin line) compared to two model spectra (thick
            lines). From one of the models we have omitted the bound-free
            cross-section of the level \ion{Mg}{v} 2p$^5$\,$^3$P$^{\rm o}$ in
            order to demonstrate that a strong autoionization feature in the
            cross-section (dashed line) can result in a line-like absorption
            feature in the calculated spectrum (marked by the arrow, \Teff=175\,000\,K).  
            }
  \label{autoionplot}
\end{figure}

\section{Model atmosphere calculations}

Line blanketed NLTE model atmospheres were computed using our {\sc PRO2} code
(Werner \etal 2003a). The models are in hydrostatic and radiative
equilibrium. The atomic models account for the most abundant elements in \hh,
carbon, oxygen and neon. Helium is also included in order to be able to derive
an upper limit for the He abundance from the \ion{He}{ii} Balmer lines in the
FUSE spectra. It is important to include Stark broadened profiles for the
strongest spectral lines (from \ion{C}{iv}, \ion{O}{vi} and \ion{Ne}{vii})
even at this stage of calculations, because they do affect the atmospheric
structure by blanketing and cooling effects (Werner 1996).

Model atmospheres with numerous parameter sets for \Teff, \logg, and He/C/O/Ne
element mixtures were computed to find the best fitting model for \hh. Based on
these model structures (keeping them fixed), we performed NLTE line formation
iterations for additional chemical elements (Mg, Al, Na, Fe, and Ni) in order
to guide our search for lines of these species in the observed spectra. Test calculations
have shown that neglecting the back-reaction of these species onto the
atmospheric structure and the emergent spectrum is unimportant for our analysis,
except for the iron group elements, which can affect the overall flux
distribution when included into the models self-consistently. Hence, we have
also computed exploratory models accounting for all elements with Z=20--28 (Ca,
Sc, Ti, V, Cr, Mn, Fe, Co, Ni), using a generic model atom comprising these
species. Table~\ref{levels_tab} summarizes all model atoms employed. For the final spectrum
synthesis fine structure splitting of atomic levels of light metals must be
considered; level populations were taken from the models with appropriate
statistical weights. For some multiplets of neon and magnesium, treatment of
the fine structure
splitting is not entirely possible because excitation energies are not known for all
sublevels. As a consequence, uncertainties of the order 0.1\AA\ in line
positions within a multiplet remain.

In the following section we describe in detail the design of our model atoms used
for the spectral analysis. This is
necessary and useful because our Chandra spectrum of \hh\ represents
the first high-resolution spectral observation of a purely photospheric stellar
spectrum in the soft X-ray region. Consequently, the majority of the detected
absorption lines were never observed before in any astrophysical object.

\begin{table}
\caption{Photospheric lines identified in the FUSE spectrum. ``*'' denotes lines
      with possibly inaccurately known wavelengths. ``:'' denotes uncertain identifications.
\label{fuse_lines_tab}
}
\begin{tabular}{llr@{\,--\,}ll}
      \hline
      \hline
%      \noalign{\smallskip}
      \noalign{\smallskip}
$\lambda$/\AA & Ion & \multicolumn{2}{c}{Transition} &  \\
      \noalign{\smallskip}
     \hline
      \noalign{\smallskip}
 948.09, 948.21   & \ion{C}{IV}   & 3s\hses           & 4p  & \\
970.55            & \ion{O}{VI}   & 5s\hses           & 8p & *\\ 
973.33            & \ion{Ne}{VII} & 2p$^1$P$^{\rm o}$ & 2p$^2$\,$^1$D  & \\
986.35            & \ion{O}{VI}   & 4s\hses           & 5p & \\
1018.11, 1018.25  & \ion{O}{VI}   & 5p                & 8d  & *\\
1031.28           & \ion{C}{IV}   & 4s\hses           & 9p  & :\\
1031.91           & \ion{O}{VI}   & 2s\hses           & 2p(3/2)  & \\
1036.67           & \ion{O}{VI}   & 5p                & 8s  & *\\
1037.08, 1037.17  & \ion{O}{VI}   & 5d                & 8f  & *\\
1037.61           & \ion{O}{VI}   & 2s\hses           & 2p(1/2)  & \\
1038.18, 1038.32  & \ion{O}{VI}   & 5f\hses           & 8d  & *\\
1038.40, 1038.42  & \ion{O}{VI}   & 5f\hses           & 8g  & *\\
1038.46, 1038.49  & \ion{O}{VI}   & 5g                & 8h  & *\\
1038.56, 1038.58  & \ion{O}{VI}   & 5g                & 8f  & *\\
1043.05, 1043.14  & \ion{O}{VI}   & 5d                & 8p  & *\\
1080.88, 1081.63  & \ion{O}{VI}   & 4p                & 5d  & \\
1097.32, 1097.34  & \ion{C}{IV}   & 4s\hses           & 8p  & :\\
1107.59, 1107.98  & \ion{C}{IV}   & 3p                & 4d  & \\
1118.25, 1118.41  & \ion{C}{IV}   & 4p                & 9s  & :\\
1122.35, 1122.62  & \ion{O}{VI}   & 4d                & 5f  & \\
1124.72, 1124.83  & \ion{O}{VI}   & 4f\hses           & 5g  & \\
1126.35, 1126.47  & \ion{O}{VI}   & 4f\hses           & 5d  & *\\
1146.79, 1147.07  & \ion{O}{VI}   & 4d                & 5p  & \\
1168.86, 1169.01  & \ion{C}{IV}   & 3d                & 4f  & \\
1171.56, 1172.44  & \ion{O}{VI}   & 4p                & 5s  & \\
1184.59, 1184.78  & \ion{C}{IV}   & 4p                & 8d  & :\\
           \noalign{\smallskip}
\hline
     \end{tabular}
\end{table}

\begin{table}
\caption{Photospheric oxygen lines identified in the Chandra spectrum. 
 ``:'' denotes uncertain identifications. \ion{O}{V} lines are not detectable
 (``-'') but included here, because their absence is discussed in the text.
\label{chandra_o_lines_tab}
}
\begin{tabular}{lr@{\,--\,}ll}
       \hline
     \hline
      \noalign{\smallskip}
$\lambda$/\AA & \multicolumn{2}{c}{Transition} & \\
      \noalign{\smallskip}
     \hline
\ion{O}{V} & & \\
     \hline
\noalign{\smallskip}
114.36 & 2s\,$^1$S & 8p\,$^1$P$^{\rm o}$ & - \\
116.16 & 2s\,$^1$S & 7p\,$^1$P$^{\rm o}$ & - \\
119.10 & 2s\,$^1$S & 6p\,$^1$P$^{\rm o}$ & - \\
124.62 & 2s\,$^1$S & 5p\,$^1$P$^{\rm o}$ & - \\
      \noalign{\smallskip}
     \hline
\ion{O}{VI} & & \\
     \hline
      \noalign{\smallskip}
93.07          & 2s\hses &      10 & :\\
93.91          & 2s\hses & \hsx  9p & :\\
95.08          & 2s\hses & \hsx  8p & \\
96.84          & 2s\hses & \hsx  7p & \\
99.60, 99.66   & 2p      &      17 & \\
99.69          & 2s\hses & \hsx 6p & \\
99.78, 99.83   & 2p      &      16 & \\
99.99, 100.00  & 2p      &      15 & \\
100.25, 100.31 & 2p      &      14 & \\
100.58, 100.63 & 2p      &      13 & \\
100.99, 101.04 & 2p      &      12 & \\
101.51, 101.57 & 2p      &      11 & \\
102.24, 102.30 & 2p      &      10 & \\
103.21, 103.26 & 2p      & \hsx  9d & \\
103.34, 103.40 & 2p      & \hsx  9s & \\
104.61, 104.67 & 2p      & \hsx  8d & \\
104.80, 104.86 & 2p      & \hsx  8s & \\
104.81         & 2s\hses & \hsx  5p & \\
106.73, 106.79 & 2p      & \hsx  7d & \\
107.02, 107.08 & 2p      & \hsx  7s & \\
110.15, 110.22 & 2p      & \hsx  6d & \\
110.66, 110.72 & 2p      & \hsx  6s & \\
115.82, 115.83 & 2s\hses & \hsx  4p & \\
116.35, 116.42 & 2p      & \hsx  5d & \\
117.33, 117.40 & 2p      & \hsx  5s & \\
129.78, 129.87 & 2p      & \hsx  4d & \\
132.22, 132.31 & 2p      & \hsx  4s & \\
150.09, 150.12 & 2s\hses & \hsx  3p & \\
\hline
     \end{tabular}
\end{table}

\begin{table}
\caption{Photospheric neon lines identified in the Chandra spectrum. 
 ``:'' denotes uncertain identifications.
\label{chandra_ne_lines_tab}
}
\begin{tabular}{lr@{\,\,--\,\,}ll}
      \hline
      \hline
      \noalign{\smallskip}
$\lambda$/\AA &  \multicolumn{2}{c}{Transition} & \\
      \noalign{\smallskip}
     \hline
\ion{Ne}{vi} & \multicolumn{2}{c}{} & \\
     \hline
      \noalign{\smallskip}
111.10, .16, .26                          &  2p\hse\,$^2$P$^{\rm o}$  & 3p\,$^2$D\hse & \\
114.07, .13, .24, .30                     &  2p\hse\,$^2$P$^{\rm o}$  & 3p\,$^2$P\hse & \\
120.21, .24, .27, .31, .35, .40, .45      &  2p$^2$\,$^4$P\hse        & 3d\,$^4$P$^{\rm o}$ & \\
121.05, .06, .10, .11, .13, .15, .19, .21 &  2p$^2$\,$^4$P\hse        & 3d\,$^4$D$^{\rm o}$ & \\
122.49, .69                               &  2p\hse\,$^2$P$^{\rm o}$  & 3d\,$^2$D\hse & \\
133.47, .48, .51, .51                     &  2p$^2$\,$^2$D\hsed       & 3d\,$^2$D$^{\rm o}$ & \\
136.22, .28, .34, .36, .37, .45, .48      &  2p$^2$\,$^4$P\hse        & 3s\hses\,$^4$P$^{\rm o}$ & \\
138.39, .64                               &  2p\hse\,$^2$P$^{\rm o}$  & 3s\hses\,$^2$S\hse & \\
     \hline
\ion{Ne}{vii} & \multicolumn{2}{c}{} & \\
     \hline
      \noalign{\smallskip}
67.88 &                                      2s$^2$\,$^1$S\hsed       & 4d\,$^1$P$^{\rm o}$ & \\
69.40 &                                      2s$^2$\,$^1$S\hsed       & 4s\,$^1$P$^{\rm o}$ & \\
74.78, .81, .87 &                            2p\hsed\,$^3$P$^{\rm o}$  & 4p\,$^3$D\hse & \\
75.76 &                                      2s$^2$\,$^1$S\hse        & 4p\,$^1$P$^{\rm o}$ & : \\ 
81.37 &                                      2p\hsw\,$^1$P$^{\rm o}$  & 4p\,$^1$P\hse & \\ 
82.16 &                                      2s$^2$\,$^1$S\hse        & 3d\,$^1$P$^{\rm o}$ & \\ 
82.17, .20, .27 &                            2p\hse\,$^3$P$^{\rm o}$  & 4d\,$^3$D\hse & \\ 
84.19, .23, .30 &                            2p\hse\,$^3$P$^{\rm o}$  & 4s\,$^3$S\hse & \\ 
85.12, .15, .22 &                            2p$^2$\,$^3$P\hse        & 4d\,$^3$D$^{\rm o}$ & \\ 
85.19, .22, .30 &                            2p$^2$\,$^3$P\hse        & 4d\,$^3$P$^{\rm o}$ & \\ 
86.47 &                                      2p$^2$\,$^1$D\hses       & 4d\,$^1$P$^{\rm o}$ & : \\ 
86.82 &                                      2p$^2$\,$^1$D\hses       & 4d\,$^1$F$^{\rm o}$ & \\ 
87.46 &                                      2s$^2$\,$^1$S\hsed       & 3s\,$^1$P$^{\rm o}$ & \\ 
87.85 &                                      2p$^2$\,$^1$D\hses       & 4d\,$^1$D$^{\rm o}$ & \\ 
89.02 &                                      2p$^2$\,$^1$D\hses       & 4s\,$^1$P$^{\rm o}$ & \\ 
89.37 &                                      2p\hsw\,$^1$P$^{\rm o}$  & 4d\,$^1$D\hse & \\ 
91.56 &                                      2p\hsw\,$^1$P$^{\rm o}$  & 4s\,$^1$S\hse & : \\ 
92.85 &                                      2p$^2$\,$^1$S\hsed       & 4d\,$^1$P$^{\rm o}$ & \\ 
94.26, .27, .30, .31, .36, .39 &             2p\hse\,$^3$P$^{\rm o}$  & 3p\,$^3$P\hse & \\ 
94.80, .84, .93 &                            2p\hse\,$^3$P$^{\rm o}$  & 3p\,$^3$S\hse & \\ 
95.65 &                                      2p$^2$\,$^1$S\hsed       & 4s\,$^1$P$^{\rm o}$ & \\ 
95.75, .81, .89, .90, .91, 96.00 &           2p\hse\,$^3$P$^{\rm o}$  & 3p\,$^3$D\hse & \\ 
97.50 &                                      2s$^2$\,$^1$S\hse        & 3p\,$^1$P$^{\rm o}$ & \\ 
103.09 &                                     2p\hsw\,$^1$P$^{\rm o}$  & 3p\,$^1$D\hse & \\ 
106.03, .08, .19 &                           2p\hse\,$^3$P$^{\rm o}$  & 3d\,$^3$D\hse & \\ 
107.10 &                                     2p\hsw\,$^1$P$^{\rm o}$  & 3p\,$^1$P\hse & \\ 
108.17 &                                     2p$^2$\,$^1$S\hsed       & 4p\,$^1$P$^{\rm o}$ & : \\ 
109.75, .78, .82, .87, .92, .98 &            2p$^2$\,$^3$P\hse        & 3d\,$^3$P$^{\rm o}$ & \\ 
110.53, .56, .59, .63, .67, .70 &            2p$^2$\,$^3$P\hse        & 3d\,$^3$D$^{\rm o}$ & \\ 
111.15 &                                     2p$^2$\,$^1$D\hses       & 3d\,$^1$P$^{\rm o}$ & \\ 
111.81 &                                     2p$^2$\,$^1$D\hses       & 3d\,$^1$F$^{\rm o}$ & \\ 
115.33, .39, .52 &                           2p\hse\,$^3$P$^{\rm o}$  & 3s\,$^3$S\hse & \\ 
115.96 &                                     2p$^2$\,$^1$D\hses       & 3d\,$^1$D$^{\rm o}$ & \\ 
116.69 &                                     2p\hsw\,$^1$P$^{\rm o}$  & 3d\,$^1$D\hse & \\ 
120.20, .27, .33, .35, .42, .48 &            2p$^2$\,$^3$P\hse        & 3s\,$^3$P$^{\rm o}$ & \\ 
121.13 &                                     2p$^2$\,$^1$D\hses       & 3s\,$^1$P$^{\rm o}$ & \\ 
121.77 &                                     2p$^2$\,$^1$S\hsed       & 3d\,$^1$P$^{\rm o}$ & \\ 
127.67 &                                     2p\hsw\,$^1$P$^{\rm o}$  & 3s\,$^1$S\hse & : \\ 
133.64 &                                     2p$^2$\,$^1$S\hsed       & 3s\,$^1$P$^{\rm o}$ & \\ 
135.29, .33, .38, .40, .50, .54 &            2p$^2$\,$^3$P\hse        & 3p\,$^3$P$^{\rm o}$ & \\ 
141.22 &                                     2p$^2$\,$^1$D\hses       & 3p\,$^1$P$^{\rm o}$ & :\\ 
     \hline
\ion{Ne}{viii} & \multicolumn{2}{c}{} & \\
     \hline
      \noalign{\smallskip}
88.08, .12 &                                 2s\hse\,$^2$S\hse        & 3p\,$^2$P$^{\rm o}$ & \\
98.11, .26 &                                 2p\hse\,$^2$P$^{\rm o}$  & 3d\,$^2$D\hse & \\
102.91, 103.08 &                             2p\hse\,$^2$P$^{\rm o}$  & 3s\,$^2$S\hse & \\
\hline
     \end{tabular}
\end{table}

\begin{table}
\caption{Photospheric magnesium lines identified in the Chandra spectrum. 
 ``:'' denotes uncertain identifications. ``bl'' denotes blends with lines of
 other elements.
\label{chandra_mg_lines_tab}
}
\begin{tabular}{lr@{\,\,--\,\,}ll}
      \hline
      \hline
      \noalign{\smallskip}
$\lambda$/\AA & \multicolumn{2}{c}{Transition} & \\
      \noalign{\smallskip}
     \hline
\ion{Mg}{v} & & \\
     \hline
      \noalign{\smallskip}
137.40, .41, .42, .74, .75, .88             & 2p$^4$\,$^3$P\hse       & 3s'\,$^3$D$^{\rm o}$ & :\\
%142.94 & 2p\,$^1$D--3s'\,$^1$D$^{\rm o}$ & \\
%146.08, .46, .62 & 2p$^4$\,$^3$P--3s\,$^3$S$^{\rm o}$ & \\
     \hline
\ion{Mg}{vi} & \multicolumn{2}{c}{} & \\
     \hline
      \noalign{\smallskip}
83.56                                       & 2p$^3$\,$^4$S$^{\rm o}$ & 4s\hse\,$^4$P & \ion{Mg}{vii} bl\\
87.40, .41                                  & 2p$^3$\,$^2$D$^{\rm o}$ & 4s\hse\,$^2$P &: \\
89.64, .65                                  & 2p$^3$\,$^2$P$^{\rm o}$ & 4s\hse\,$^2$P &: \\
93.52                                       & 2p$^3$\,$^2$D$^{\rm o}$ & 3d''\,$^2$D &: \\
95.38, .42, .48                             & 2p$^3$\,$^4$S$^{\rm o}$ & 3d\hse\,$^4$P & \ion{Mg}{vii} bl\\
96.08, .09                                  & 2p$^3$\,$^2$P$^{\rm o}$ & 3d''\,$^2$D & \\
96.26, .30                                  & 2p$^3$\,$^2$D$^{\rm o}$ & 3d'\,$^2$P & \\
97.25, .28                                  & 2p$^3$\,$^2$D$^{\rm o}$ & 3d'\,$^2$F & \\
98.50, .51                                  & 2p$^3$\,$^2$P$^{\rm o}$ & 3d'\,$^2$S & \\
98.98, .99, 99.02, .04                      & 2p$^3$\,$^2$P$^{\rm o}$ & 3d'\,$^2$P & \\
100.70, .90                                 & 2p$^3$\,$^2$D$^{\rm o}$ & 3d\,$^2$F & \\
101.49, .55                                 & 2p$^3$\,$^2$D$^{\rm o}$ & 3d\,$^2$P & \ion{O}{vi} bl\\
104.52, .53, .59, .60                       & 2p$^3$\,$^2$P$^{\rm o}$ & 3d\hse\,$^2$P &  \ion{O}{vi} bl\\
111.17, .19                                 & 2p$^3$\,$^2$P$^{\rm o}$ & 3s''\,$^2$S &  \ion{Ne}{vii} bl\\
111.55, 75, .86                             & 2p$^3$\,$^4$S$^{\rm o}$ & 3s\hse\,$^4$P & :\\
113.19                                      & 2p$^3$\,$^2$D$^{\rm o}$ & 3s'\,$^2$D & \\
116.97, 117.22                              & 2p$^3$\,$^2$D$^{\rm o}$ & 3s\hse\,$^2$P & \\
116.97, .99                                 & 2p$^3$\,$^2$P$^{\rm o}$ & 3s'\,$^2$D & \\
117.55                                      & 2p$^4$\,$^2$D\hse       & 3s$^{\rm v}$\,$^2$P$^{\rm o}$ & \\
121.01, .03, .29, .30                       & 2p$^3$\,$^2$P$^{\rm o}$ & 3s\hse\,$^2$P & \ion{Ne}{vii} bl\\
123.59                                      & 2p$^4$\,$^2$D\hse       & 3s$^{\rm iv}$\,$^2$D$^{\rm o}$ :& \\
130.31, .64                                 & 2p$^4$\,$^2$P\hse       & 3s$^{\rm v}$\,$^2$P$^{\rm o}$ & \\
137.81, 138.17                              & 2p$^4$\,$^2$P\hse       & 3s$^{\rm iv}$\,$^2$D$^{\rm o}$ :& \\
\hline
    \end{tabular}
\end{table}

\addtocounter{table}{-1}
\begin{table}
\caption{continued}
\begin{tabular}{lr@{\,\,--\,\,}ll}
      \hline
      \hline
      \noalign{\smallskip}
$\lambda$/\AA & \multicolumn{2}{c}{Transition} & \\
      \noalign{\smallskip}
\hline
\ion{Mg}{vii} &\multicolumn{2}{c}{} & \\
     \hline
      \noalign{\smallskip}
78.34, .41, .52                  & 2p$^3$\,$^3$P\hse         & 3p\hse\,$^3$P$^{\rm o}$ & \\
80.95, 81.02, .14                & 2p$^3$\,$^3$P\hse         & 3p\hse\,$^3$S$^{\rm o}$ & :\\
82.94, 82.97, 83.01              & 2p$^3$\,$^5$S$^{\rm o}$   & 3d\hse\,$^5$P\hse & \\
83.51, .56, .59, .64, .71, .76   & 2p$^3$\,$^3$P\hse         & 3d\hse\,$^3$P$^{\rm o}$ & \\
83.91, .96, .99, 84.02, .09, .11 & 2p$^3$\,$^3$P\hse         & 3d\hse\,$^3$D$^{\rm o}$ & \\
85.34                            & 2p$^2$\,$^1$D\hses        & 3d\hse\,$^1$P$^{\rm o}$ & :\\
85.41                            & 2p$^2$\,$^1$D\hses        & 3d\hse\,$^1$F$^{\rm o}$ & \\
87.72                            & 2p$^2$\,$^1$D\hses        & 3d\hse\,$^1$D$^{\rm o}$ & \\
87.89                            & 2p$^2$\,$^1$D\hses        & 3d\hse\,$^3$F$^{\rm o}$ & intercomb.\\
88.68                            & 2p$^2$\,$^1$S\hsed        & 3d\hse\,$^1$P$^{\rm o}$ & \\
94.04, .17, .24                  & 2p$^3$\,$^5$S$^{\rm o}$   & 3s\hse\,$^5$P\hse & \ion{Mg}{viii} bl \\
95.03, .04                       & 2p$^3$\,$^3$D$^{\rm o}$   & 3s'\,$^3$D & \\
95.26, .38, .42, .49, .56, .65   & 2p$^3$\,$^3$P\hse         & 3s\hse\,$^3$P$^{\rm o}$ & \\
98.03                            & 2p$^2$\,$^1$D\hses        & 3s\hse\,$^1$P$^{\rm o}$ & \\
98.98                            & 2p$^3$\,$^3$P$^{\rm o}$   & 3s'\,$^3$D\hse & \\
101.96, .97, 102.14, .23         & 2p$^3$\,$^3$D$^{\rm o}$   & 3s\hse\,$^3$P\hse & \\
102.47                           & 2p$^2$\,$^1$S\hsed        & 3s\,$^1$P$^{\rm o}$ & \\
105.17                           & 2p$^3$\,$^1$D$^{\rm o}$   & 3s'\,$^1$D\hse & \\
106.52, .71, .81                 & 2p$^3$\,$^3$P$^{\rm o}$   & 3s\hse\,$^3$P\hse & \ion{O}{vi} bl \\
110.11                           & 2p$^3$\,$^1$P$^{\rm o}$   & 3s'\,$^1$D\hse & \ion{O}{vi} bl\\
111.98, 112.00, .11, .12, .27    & 2p$^3$\,$^3$D$^{\rm o}$   & 3p\hse\,$^3$P\hse & \\
117.43, .66, .78                 & 2p$^3$\,$^3$S$^{\rm o}$   & 3s\hse\,$^3$P\hse & \\
117.52, .64, .81                 & 2p$^3$\,$^3$P$^{\rm o}$   & 3p\hse\,$^3$P\hse & \\
130.94, 131.09, .30              & 2p$^3$\,$^3$S$^{\rm o}$   & 3p\hse\,$^3$P\hse & :\\
%144.30, .74, .92                 & 2p$^4$\,$^3$P--3p\,$^3$S$^{\rm o}$ & \\
\hline
\ion{Mg}{viii} &\multicolumn{2}{c}{}  & \\
     \hline
      \noalign{\smallskip}
69.41, .47, .57                  & 2p\hse\,$^2$P$^{\rm o}$   & 3p\hse\,$^2$D\hse & \\
74.27, .32, .34, .37, .41, .43   & 2p$^2$\,$^4$P\hse         & 3d\hse\,$^4$D$^{\rm o}$ & \\
74.86, 75.03, .04                & 2p\hse\,$^2$P$^{\rm o}$   & 3d\hse\,$^2$D\hse& \\
80.23, .25                       & 2p$^2$\,$^2$D\hses        & 3d\hse\,$^2$D$^{\rm o}$ & \\
81.73, .79, .84, .87, .94, .98   & 2p$^2$\,$^4$P\hse         & 3s\hse\,$^4$P$^{\rm o}$ & \\
82.60, .82                       & 2p\hse\,$^2$P$^{\rm o}$   & 3s\hse\,$^2$S\hse & \\
86.24, .36, .38                  & 2p$^2$\,$^2$P\hse         & 3d\hse\,$^2$D$^{\rm o}$ & \\
86.84, .85, 87.02                & 2p$^2$\,$^2$D\hses        & 3s\hse\,$^2$P$^{\rm o}$ & :\\
92.13, .32                       & 2p$^2$\,$^2$S\hsed        & 3s\hse\,$^2$P$^{\rm o}$ & :\\
93.89, 94.07, .10, .27           & 2p\hse\,$^2$P\hse         & 3s\hse\,$^2$P$^{\rm o}$ & :\\
98.13, .15                       & 2p$^3$\,$^4$S$^{\rm o}$   & 3p\hse\,$^4$P\hse & \ion{Ne}{viii} bl\\
102.13, .21                      & 2p$^2$\,$^2$S\hsed        & 3p\hse\,$^2$P$^{\rm o}$ & :\\
105.97, .98, 106.0               & 2p$^3$\,$^2$D$^{\rm o}$   & 3p\hse\,$^2$P\hse &  \ion{Ne}{vii} bl\\
106.81, .83                      & 2p$^3$\,$^2$P$^{\rm o}$   & 3p\hse\,$^2$S\hse & \ion{O}{vi} bl\\
108.93, 109.18, .20              & 2p$^3$\,$^2$P$^{\rm o}$   & 3p\hse\,$^2$D\hse & \\
114.90, .91, .93, .94            & 2p\hse\,$^2$D$^{\rm o}$   & 3d\hse\,$^2$D\hse & :\\
\hline
    \end{tabular}
\end{table}

\begin{figure*}[tbp]
\includegraphics[width=16cm]{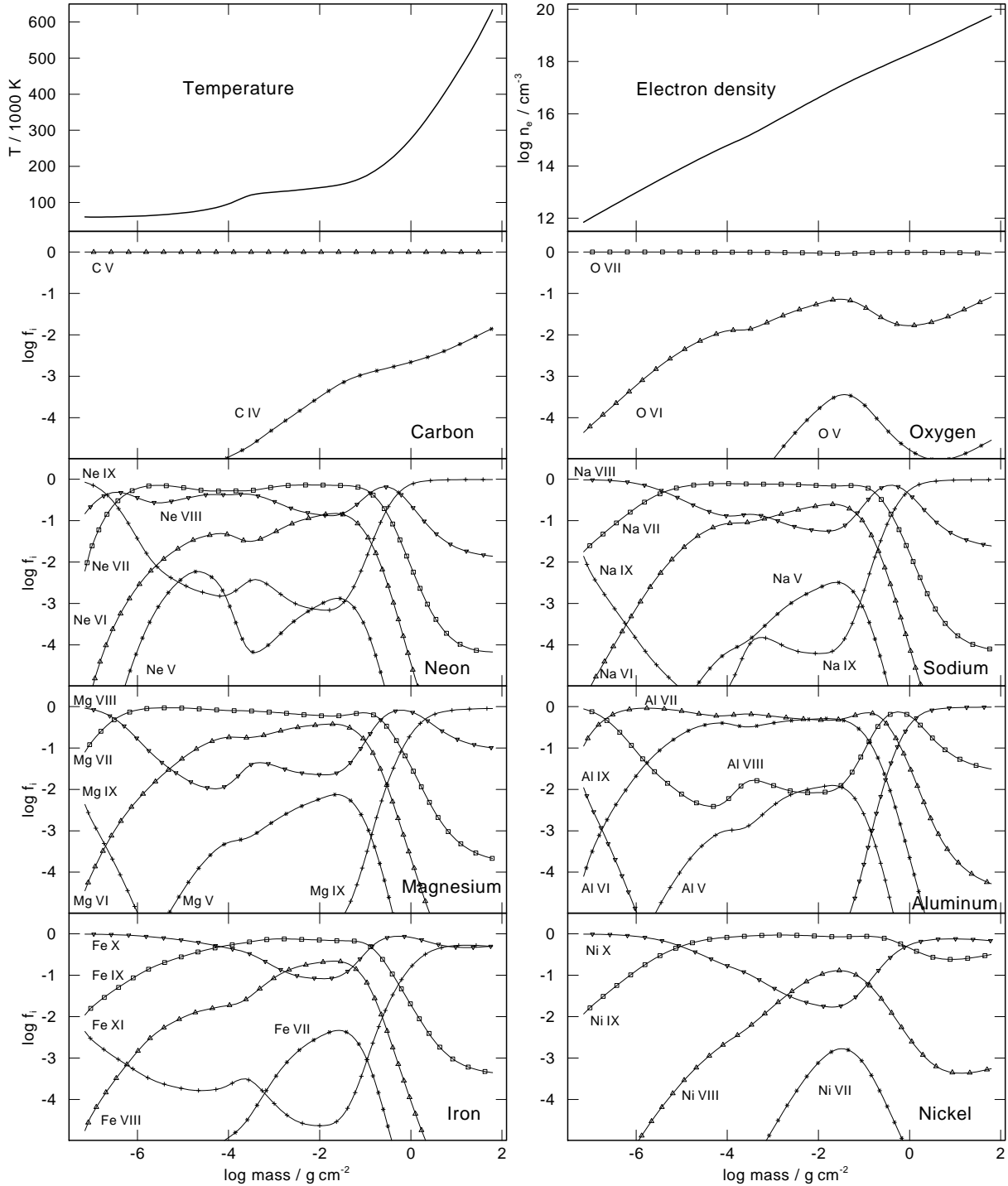}
  \caption[]{
Depth dependence of temperature, electron density, and ionization fractions of
            chemical species in the model with \Teff=175\,000\,K.
            }
  \label{ionplot}
\end{figure*}

\section{Atomic data}

We have performed a detailed search for reliable atomic data in a number of
atomic data sources. These are the following databases:

\noindent
(i) Kurucz (1991) level and line lists\footnote{http://cfa-www.harvard.edu/amdata/ampdata/kurucz23/sekur.html},\\
(ii) Opacity Project (OP, Seaton \etal 1994) TOPbase\footnote{http://legacy.gsfc.nasa.gov/topbase/home.html},\\ 
(iii) National Institute of Standards and Technology (NIST)\footnote{http://physics.nist.gov/},\\
(iv) {\sc Chianti} database (Young \etal 2003)\footnote{http://wwwsolar.nrl.navy.mil/chianti.html},\\
(v) Kelly Atomic Line Database\footnote{http://cfa-www.harvard.edu/amdata/ampdata/kelly/kelly.html},\\
(vi) University of Kentucky Atomic Line List\footnote{http://www.pa.uky.edu/$^\sim$peter/atomic/},

\noindent
as well as the monograph of Bashkin \& Stoner (1975). The Kurucz and OP
databases are our basic sources for atomic data of heavy and light metals,
respectively. Bashkin \& Stoner is an important source for accurate level
energies and  line positions for light metals. Important additional data in this
respect were found in the NIST and {\sc Chianti} databases.

For the light metals (C-Na) the most serious problem in many cases is the lack
of accurate line positions for highly ionized species, often making individual
line identifications difficult or impossible.  The same problem is faced with Ca
and the iron group elements: Their combined opacity is important for the overall
flux distribution, but identifying these species by means of individual lines is
hardly possible. However, it turns out that the cumbersome work for a critical
data compilation can yield satisfying results. A large number of multiplets from
Ne and Mg can be identified in the Chandra data.

Line broadening is a critical issue and introduces considerable uncertainties
for detailed line-profile fits. Linear Stark broadening is assumed for
ions with one valence electron (\ion{He}{ii}, \ion{C}{iv}, \ion{O}{vi}) and treated in an
approximate way, while quadratic Stark broadening is assumed for all other ions
(for details see Werner \etal 1991).

Another problem is posed by the requirement of good photoionization (b-f)
cross-sections for metals. The strengths of absorption edges within the Chandra
spectral range and the adjacent EUV/FUV region affects the spectral energy
distribution. We have investigated this problem, as well as the effect of
pressure ionization on the absorption edges, in detail before and refer to
our earlier EUVE analysis of \hh\ ({\small WW99}). What is a new problem,
that becomes important with the superior Chandra spectral resolution, is the
occurrence of autoionization resonances in the b-f cross-sections. Some of these
resonances are strong and narrow, mimicking absorption lines in the spectra
(Fig.\,\ref{autoionplot}). Although these resonances are included in the OP
data, their wavelength position is too uncertain for a unique identification in
the observations. As a consequence, hitherto unidentified lines in the Chandra
spectrum may be due to \mbox{b-f} resonances of metals and, vice versa, absorption
features in the synthetic spectra without an observed counterpart could be due
to such resonances in the models. For this reason we have substituted the OP b-f
data by hydrogen-like cross-sections in the detailed final spectral fits shown
in this paper. This does change the continuum flux slightly, but
in view of the difficulties we encounter with fitting the overall flux
distribution (see below), this is considered unimportant. Generally, strong
resonances are fortunately not very frequent and can be identified in the models
on a case-by-case basis.

Let us now present the design of our model atoms. A good
orientation concerning the ionization degree of different species is provided
by Fig.\,\ref{ionplot}. Besides the temperature and electron density
stratification of a model with \Teff=175\,000\,K it shows the ionization
fractions of all considered chemical species. We will discuss the changes in
ionization balance with \Teff\ in Sect.\,\ref{sect_5}.

\subsection{Helium, carbon, and oxygen}

Because of the high temperatures in the atmosphere of \hh, 
helium is strongly ionized so that a one-level
\ion{He}{i} model atom is sufficient.

For C and O, the helium-like noble-gas configurations (\ion{C}{v}, \ion{O}{vii})
represent the highest ionization stages in the model atoms. They are also the
dominant ionization stages in our model atmospheres (see
Fig.\,\ref{ionplot}). Test calculations based upon extended model atoms including
H-like ions \ion{C}{vi}, \ion{O}{viii} showed that this is sufficient. The
population of the latter ions becomes significant only in the outermost and
innermost layers of the hottest models calculated (\Teff=250\,000\,K). This 
neither changes the observed \ion{C}{iv} and \ion{O}{vi}  line profiles nor the
flux distribution in the wavelength bands studied here.  Also, the population of
excited levels in \ion{C}{v} and \ion{O}{vii} is very weak, which explains why
no subordinate lines from these ions are detected in the observed data (the
respective resonance lines are located blueward of the recorded Chandra
spectrum).

The UV/optical spectrum of \hh\ is dominated by \ion{C}{iv} and \ion{O}{vi}
lines (but note that no \ion{C}{iv} lines are located in the Chandra spectral
range). Available level energies often yield line positions that do not coincide
with line positions from laboratory measurements. This is frequently the case
for \ion{O}{vi} lines, but also sometimes for \ion{C}{iv} lines. When an exact line
position could be taken from the databases, we shifted the synthetic line to that position,
but uncertainties remain for many lines, particularly in the FUV range. Similar
problems occur with many \ion{O}{v} lines in the Chandra range, preventing their
identification, which is unfortunate because these lines would be sensitive
\Teff\ indicators. In all cases we restricted detailed profile analyses to
lines with accurately known wavelengths. Tables \ref{fuse_lines_tab} and \ref{chandra_o_lines_tab} list the
identified C and O lines.

\subsection{Neon, magnesium, aluminum, and sodium}

Besides \ion{O}{vi} lines, lines from \ion{Ne}{vii} are the strongest absorption
features in the Chandra spectrum of \hh. This ion also has a strong optical line
at 3644\AA, which we detected in a Keck spectrum ({\small WW99}), and in the present paper we
announce the detection of the \ion{Ne}{vii}~973.3\AA\ absorption line 2p$^1$P$^{\rm
o}$--2p$^2$\,$^1$D; to our knowledge it is the first time ever seen in an astrophysical spectrum.  There
is some uncertainty in the literature concerning the accurate wavelength
position of this line. The first measurement was published by Johnston \& Kunze
(1971) who give 973.6\AA. Lang (1983) quotes 973.33\AA\
(without reference) and this value is also found in the Chianti
database. According to Kramida (NIST, priv.\ comm.) the best measurement for
this line was done by Lindeberg (1972). The measured wavelength is
973.302$\pm$0.005\AA. For our synthetic spectra we have adopted
973.33\AA.  We also model the line spectrum of \ion{Ne}{vi} and
\ion{Ne}{viii} and successfully identify these ions in the Chandra data (see
below). Table\,\ref{chandra_ne_lines_tab} gives a complete list of all lines
considered in the spectrum synthesis calculations. As already mentioned, fine
structure splitting of multiplets is problematic in those cases, in which
knowledge of energy levels within one term is incomplete. This can cause
uncertainties in wavelength positions of individual lines within a multiplet of
the order 0.1\AA. We have omitted all lines involving energy levels uncertain to
an extent that the corresponding wavelength uncertainty is much larger than
0.1--0.2\AA. As a consequence, many unidentified lines in the Chandra spectrum
probably stem from such Ne lines.

After first test calculations with a small magnesium model atom we found that some lines
of these species might be present in the Chandra data. However, because the Mg
lines are much weaker than O and Ne lines, 
we have designed a very detailed model atom for Mg in order to look for as many
lines of this species as possible. After a critical data compilation and the
rejection of many lines because of wavelength uncertainties there is
no doubt that we can identify many lines of \ion{Mg}{v-viii}. 
As in the case of Ne, many unidentified Chandra spectral lines
certainly stem from Mg lines with uncertain wavelength positions. It is
interesting to note that we predict an observable intercombination line, which
indeed is identified in the Chandra data (Table~\ref{chandra_mg_lines_tab}).

In the very same way we tried to identify Al and Na lines in the Chandra
spectrum. Although large model atoms were constructed, we cannot detect these
species beyond doubt. Details of this search will be described below.

\begin{figure*}[tbp]
\includegraphics[width=\textwidth]{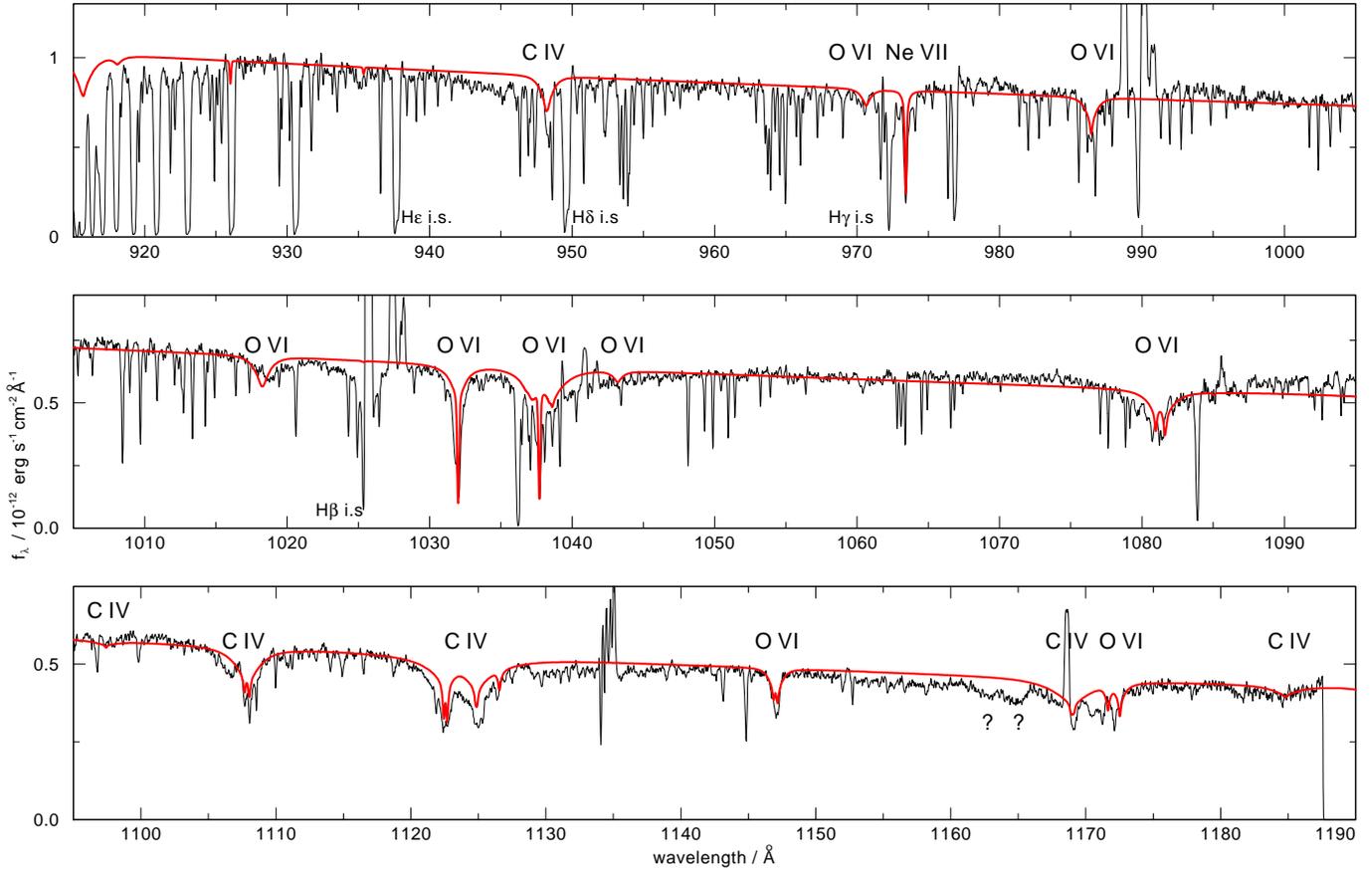}
  \caption[]{
The FUSE spectrum of \hh\ with the strongest photospheric lines identified. They
            stem from \ion{C}{iv} and \ion{O}{vi}. Note particularly the strong
            \ion{Ne}{vii} line at 973.3\AA. This line is identified for the very
            first time in a stellar spectrum. All emission features are
            geocoronal, not photospheric. Overplotted is a model spectrum
            (\Teff=200\,000\,K). Observed and model spectra are convolved with Gaussians
            with FWHM=0.05\AA\ and 0.1\AA, respectively.  
            }
  \label{fuse_overview}
\end{figure*}

\begin{figure*}[tbp]
  \resizebox{\hsize}{!}{\includegraphics{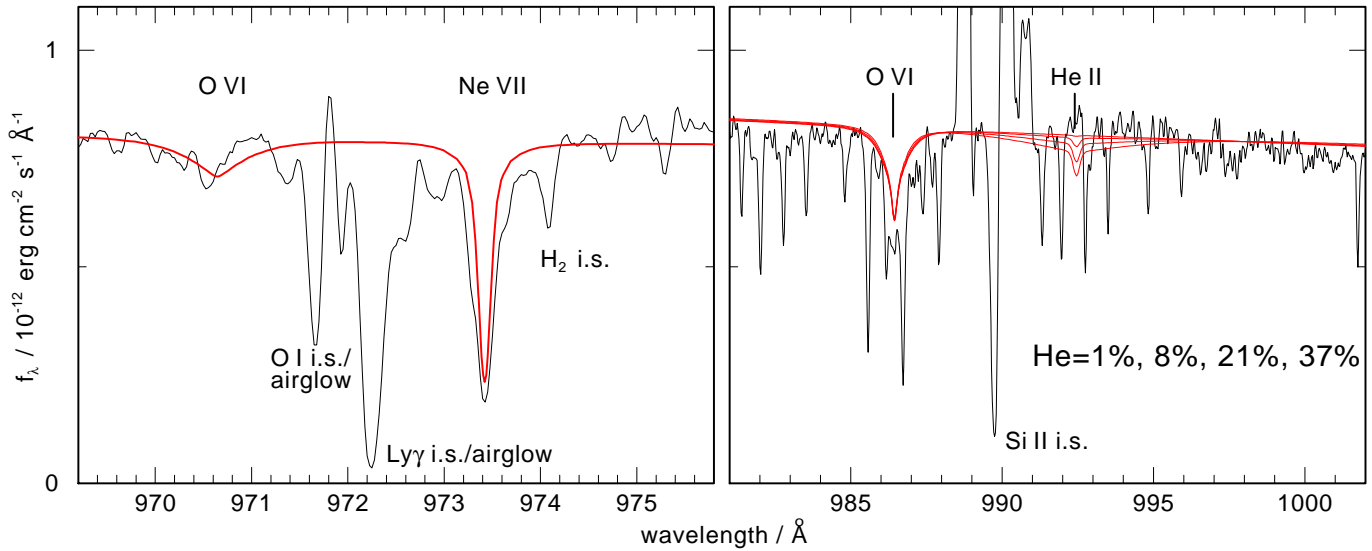}}
  \caption[]{
Details from the FUSE spectrum. Left panel: region around the newly identified \ion{Ne}{vii}
            line. Right panel: Close-up of the region near the \ion{He}{ii}
            (n=2$\rightarrow$7) line. It is not detectable. Overplotted are four
            models with increasing He abundance, confirming the He-deficiency of
            \hh\ (\Teff=200\,000\,K). Observed and model spectra are convolved with Gaussians
            with FWHM=0.05\AA\ and 0.1\AA, respectively.
            }
  \label{fuse_detail}
\end{figure*}

\subsection{Calcium and iron group elements}

Synthetic spectra including lines from Ca and Fe group elements were calculated
using a superlevel/superline approach which is described, e.g., in Rauch \&
Deetjen (2003). All line data were taken from Kurucz's line lists, except for
\ion{Fe}{x}, which were taken from the Opacity Project's TOPbase. \ion{Ni}{x}
lines exist neither in those two datasets nor in the NIST database and, hence,
could not be considered here. Kurucz's data
compilation consists of two sets. The first one includes only lines with
accurate wavelengths
determined from laboratory experiments (``POS'' data). The second list
additionally includes predicted lines with uncertain wavelengths (``LIN''
data). Our strategy is that we perform model and line formation iterations using the
full (LIN) line list, because precise wavelength positions are unimportant at this
stage of calculations, but accounting for the complete line opacities is important. The final
synthetic spectra that are compared with observations, however, are computed
using the POS data only, in order to facilitate line identifications. We shall
see that strong blanketing by the large number of LIN lines makes the identification of
individual POS lines of iron and nickel
virtually impossible. The OP
\ion{Fe}{x} lines are treated like Kurucz's LIN lines, because their
positions are uncertain, too.

\section{Overview of FUSE and Chandra Observations}
\label{obs}

\hh\ was observed by FUSE on four separate occasions, with a 
total integration time of 67 ksec.  The FUSE instrument consists of four
coaligned telescopes, each with a prime-focus spectrograph.  
The observation log is given in Table~\ref{tab_fuselog}; all data were 
obtained in time-tag mode and processed with CALFUSE v2.4.  
The LWRS spectrograph apertures were used for all observations, hence the
zero-point of the wavelength scale is uncertain to within about $\pm 0.15$\AA. 
All exposures were photometric, or nearly so, in all channels.   
Descriptions of the FUSE instrument, and channel alignment
and wavelength calibration issues are given by Moos \etal (2000) and
Sahnow \etal (2000).  The wavelength
scales for spectra from the individual exposures were coaligned separately for
each channel by means of narrow interstellar absorption lines.  Spectra from
channels other than LiF1 were then given an additional zero-point offset so
that ${\rm H_{2}}$ lines (which are present in all channels) had a common
heliocentric velocity.  The spectra from all the channels were then resampled
onto a common wavelength scale with 0.025\AA\ pixels, and 
combined.  The resulting spectrum covers 912 - 1187\AA\ with a typical
spectral resolution of 0.08--0.10\AA.

\begin{table}
%\begin{center}
\caption{Log of FUSE observations of \hh. \label{tab_fuselog}}
\begin{tabular}{l c r}
\hline
\hline
Prog ID & Date & ${\rm T_{exp}}$(ksec) \\
\hline
Q1090202  &  2000-02-19 & 11.3 \\
S6010202  &  2002-01-28 & 32.7 \\
M1052601  &  2003-01-13 & 13.9 \\
M1052602  &  2003-05-06 &  8.9 \\
\hline
\end{tabular}
%\end{center}
\end{table}

The spectrum is characterized by a few broad photospheric lines of
C\,IV and O\,VI and many narrow interstellar lines, predominantly from H$_2$
(see Fig.\,\ref{fuse_overview}). The identified photospheric lines are listed in
Table~\ref{fuse_lines_tab}. All other absorption features are of interstellar
origin, except for two broad and shallow features at 1163\AA\ and 1165\AA\
(marked by ``?'' in Fig.\,\ref{fuse_overview}), which we think are of
photospheric origin,
but cannot be identified. They are also seen, even more prominently, in other
very hot PG1159 stars, the central stars \object{K1-16} and
\object{RX\,J2117+3412}. 
There are \ion{S}{viii} lines listed at these wavelengths, however, this
identification is regarded as unlikely, because of the high level energies and because other lines from this ion should
be detectable, too, which is not the case. The model spectra shown here are
shifted by +0.09\AA\ in order to fit the bulk of the observed lines.

Two interesting details are shown in
Fig.\,\ref{fuse_detail}. The left panel focuses on the region around a rather
strong absorption line, which stems from \ion{Ne}{vii}. As already mentioned
this is the first detection of this line in a stellar spectrum. It is also
seen in FUSE spectra of other PG1159 stars (e.g.\ in the prototype \object{PG1159-035} itself). The overplotted
model, which fits the observed line very well, has a Ne abundance of 2\%, in
agreement with the results from the Chandra spectrum analysis described below.  The right
panel of Fig.\,\ref{fuse_detail} is centered around the location of a \ion{He}{ii}
line. \ion{He}{ii} absorption is not detectable and a comparison with models shows that \hh\
is strongly helium-deficient. This corroborates the result of an optical
spectral analysis (He$<$1\%), which was based on the lack of a
He\,II~4686\AA\ emission line ({\small W91}). From 
Fig.\,\ref{fuse_detail} we derive a (less tight) upper limit for the He
abundance, namely He$<$5\%.

\begin{figure*}[tbp]
  \resizebox{\hsize}{!}{\includegraphics{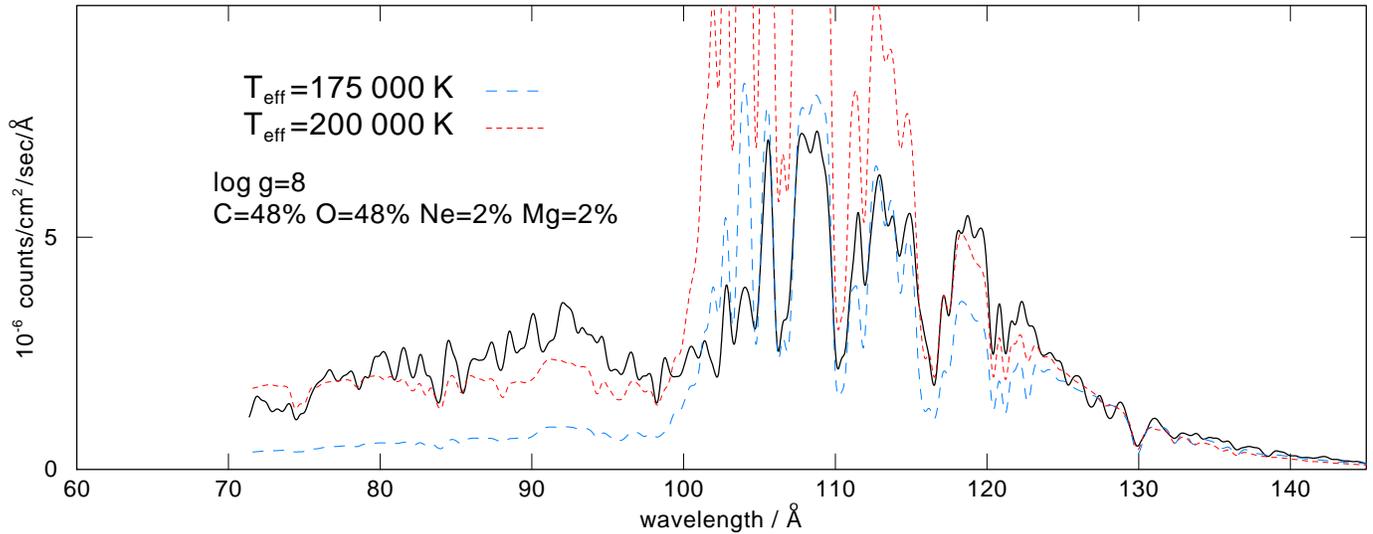}}
  \caption[]{
Overview of the Chandra spectrum of \hh\ (solid line). Two models with
different $T_{\rm eff}$ are shown. The 175\,000\,K model (long dashed line) fits the overall flux
at long wavelengths, while it underestimates the short wavelength flux. 
The hotter model (200\,000\,K, short dashed line) fits at short wavelengths but
overestimates the long wavelength flux. The model spectra were attenuated by an
ISM model with $n_{\rm H}=9.02 \cdot 10^{19}$\,cm$^{-2}$ and $1.06 \cdot
10^{20}$\,cm$^{-2}$, respectively (assuming $n_{\rm He}/n_{\rm H}=0.1$, and 
$n_{\rm \ion{He}{i}}/n_{\rm \ion{He}{ii}}=0.5$), then folded through the instrument response
and normalized to the observation to fit near 130\AA. The models include He, C,
O, Ne, Mg as described in the text. For clarity, all spectra
are smoothed with Gaussians (0.5\AA\ FWHM).
            }
  \label{chandra_overview}
\end{figure*}

\hh\ was observed with Chandra on Sep.\ 27, 2000, with an integration time of 7
hours. Flux was detected in the range 60\AA--160\AA\ and the spectral resolution
is about 0.1\AA. Fig.\,\ref{chandra_overview} shows the overall spectrum. It is
characterized by a roll-off at long wavelengths due to ISM absorption. The
maximum flux is detected near 110\AA. Between 105\AA\ and 100\AA\ the flux drops
because of photospheric absorption from the \ion{O}{vi} edge caused by the first excited atomic level. The
edge is not sharp because of a converging line series and pressure ionization
(see {\small WW99} for detailed model spectra). Below 100\AA\ the flux
decreases, representing the Wien tail of the photospheric flux
distribution. Fig.\,\ref{chandra_overview} demonstrates that our models cannot
fit the entire wavelength range at a unique temperature, however, from the
overall flux shape $T_{\rm eff}$ can be constrained between 175\,000\,K and
200\,000\,K.

We have looked for possible explanations for the inability of our models to fit
the overall X-ray flux distribution. We must look for processes which decrease
the flux level at $\lambda>100$\AA\ and/or increase the flux level at shorter
wavelengths.  Line blanketing by heavy metals (introduced below) does not solve
the problem. Including the large number of absorption lines blocks flux at
$\lambda>100$\AA, but this flux re-emerges in the same wavelength region just in
between spectral windows of strong line blanketing, hence, a flux increase at
$\lambda<100$\AA\ cannot be enforced by including this opacity source. From this result we
feel that we must look for a missing continuous opacity.

One possibility was that our carbon model atom needed to be extended to higher
ionization stages. Excited levels of \ion{C}{v} (2s and 2p) potentially cause
absorption edges between 130--150\AA. This could decrease the flux for
$\lambda>100$\AA\ and at the same time, because of total flux constancy, could increase
the flux $\lambda<100$\AA. However, a test calculation has shown that this does not
solve the problem. Even in a  model with \Teff=250\,000\,K, in which \ion{C}{v}
is the dominant ionization stage of carbon in most parts of the atmosphere, only
weak \ion{C}{v} edges appear, decreasing the maximum model flux near 110\AA\ by
less than 10\%. By analogy, excited  \ion{O}{vii} levels can cause absorption
edges. But their strength is expected to be even less important, because the
ionization potential of \ion{O}{vi} is higher than that of \ion{C}{iv}.

We also tried to vary other model parameters to overcome this
problem. Increasing the surface gravity from \logg\,=\,8 to \logg\,=\,9 causes a flux
decrease of about 50\% in the 100--130\AA\ range. The additional opacity is
caused by \ion{Ne}{vii} and \ion{O}{vi} line wings which become very strongly
pressure broadened. In addition, increased pressure ionization of \ion{O}{vi}
causes a shift of the strong ionization edges towards lower wavelengths. But
again, the result is not a flux increase at $\lambda<100$\AA, but a higher flux at
$\lambda>130$\AA. A similar result was obtained from an exercise, where we increased
the (uncertain) line broadening parameters even by very large amounts.  The
oxygen abundance is also not a critical parameter. Tests show that the strength of the
\ion{O}{vi} absorption edge near 100\AA\ is insensitive to variations of the O
abundance within reasonable limits. Even a reduction of the O abundance by a factor
of 10 decreases the continuum flux jump at the absorption edge from 95\% only to 80\%. In the same direction, we artificially
decreased the b-f cross-section of the \ion{O}{vi} levels by factors larger than
we expect from the uncertainty in the atomic data for this relatively simple
one-electron ion. Switching between OP
cross-sections and hydrogen-like cross-sections has only a weak effect on the
model flux.

To summarize these tests, the only way to significantly increase the short
wavelength flux is to increase the effective temperature to roughly
200\,000\,K. An unknown opacity source must be responsible for suppressing
the excess model flux at longer wavelengths. A detailed analysis of the Chandra
line spectrum can be used to further constrain the effective temperature by using
ionization balances of several species. We will now show that this supports
the \Teff=200\,000\,K estimate.

\begin{figure*}[tbp]
\includegraphics[width=17.1cm]{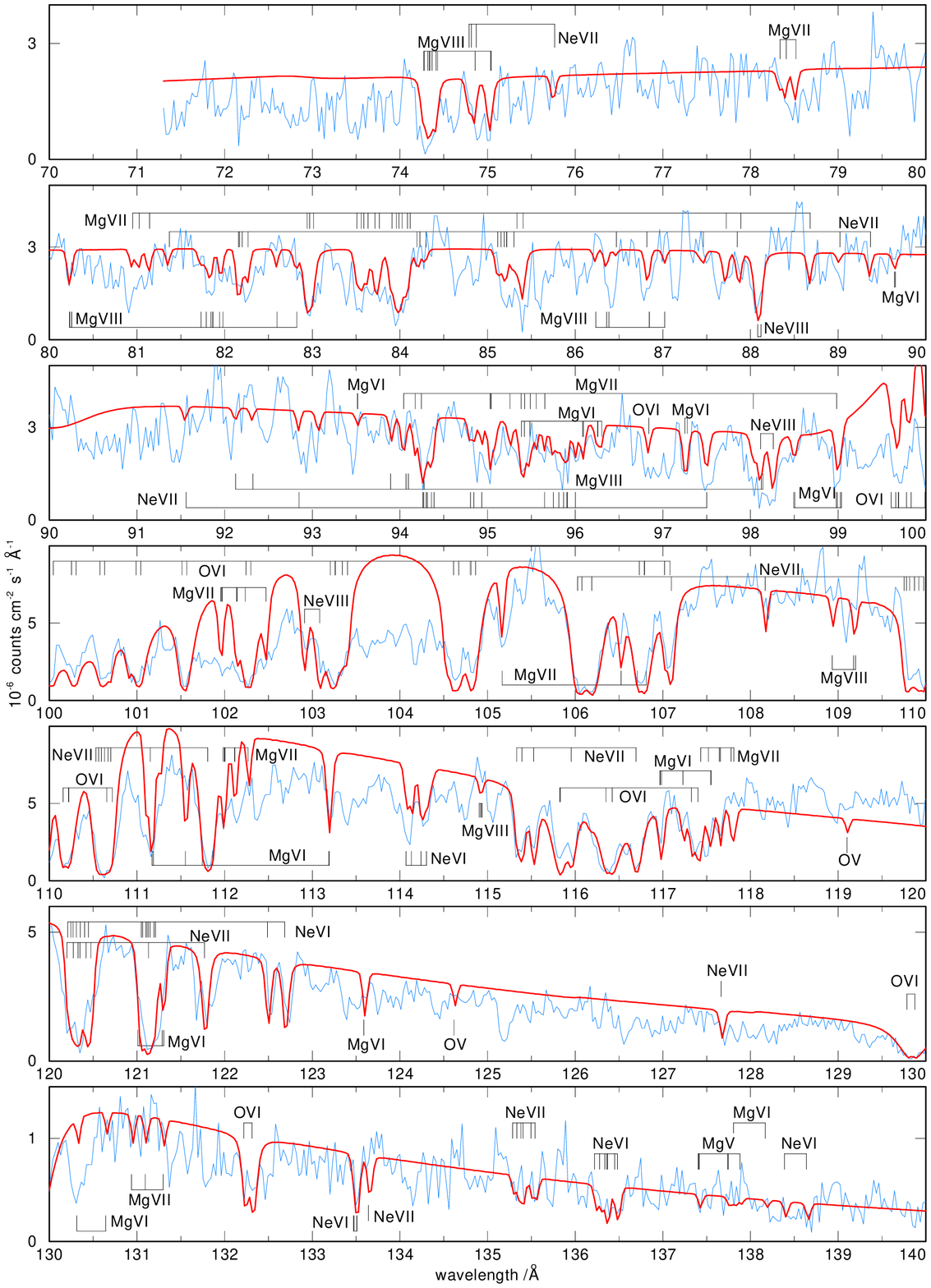}
  \caption[]{
Expanded view of the Chandra spectrum and a model with
            \Teff=200\,000\,K (including He, C, O, Ne, and Mg as described in the
            text). 
            }
  \label{chandra_fit}
\end{figure*}

\section{Analysis of the X-ray absorption line spectrum}\label{sect_5}

Figure~\ref{chandra_fit} shows in more detail the Chandra spectrum of
\hh. Considering the problems with fitting the overall continuum flux, we now
normalize the model flux to the local continuum in each panel of this figure. In
order to facilitate line identifications, we overplot that model from our grid
which turned out to fit best to the line spectrum. The observed spectrum was 
shifted by -0.06\AA\ to the rest wavelength. We detect lines from
\ion{O}{vi}, \ion{Ne}{vi-viii}, and \ion{Mg}{v-viii}. The strong and broad
\ion{O}{vi} and \ion{Ne}{vii} lines were already detected in the EUVE spectrum
({\small WW99}), the detection of magnesium is new. Identification of even
the strongest lines becomes difficult at longer wavelengths (see bottom panel of
Fig.\,~\ref{chandra_fit}) due to the decreased S/N-ratio.

For a concise presentation of our analysis, we first concentrate on the effective
temperature which is the most important parameter determining ionization
balances, while keeping fixed \logg\,=\,8 from our analysis of optical spectra. We
think that an improvement of the error range for the gravity ($\pm$0.5\,dex, {\small W91}) is
not possible because of uncertainties in line broadening calculations. We also
keep fixed the relative abundances of C, O, and Ne, which are the dominant
species, and include He as a trace element (which has unimportant effects on
model structures and fluxes). Hence, the composition of all models shown here
is: He=1\%, C=48.5\%, O=48.5\%, Ne=2\%. As mentioned above, other elements are included
later in line formation iterations.

We start the \Teff\ analysis with a first estimate for the Mg abundance
(2\%) which will be justified later. After this we turn to the search for other
species (Al, Na, Fe, Ni).

\subsection{Effective temperature from ionization balances}

Given the described uncertainty in the determination of \Teff\ from the overall X-ray flux
shape, it is important to constrain the temperature by comparing line strengths
from different ionization stages of the identified species. We discuss their
appearance in the context of three models with \Teff=175\,000\,K, 200\,000\,K, and
250\,000\,K. Qualitatively, the shift of ionization balances and, hence, the
change of line strengths with increasing \Teff\ can be understood with the help
of Fig.\,\ref{ionplot}.

\begin{figure*}[tbp]
  \resizebox{\hsize}{!}{\includegraphics{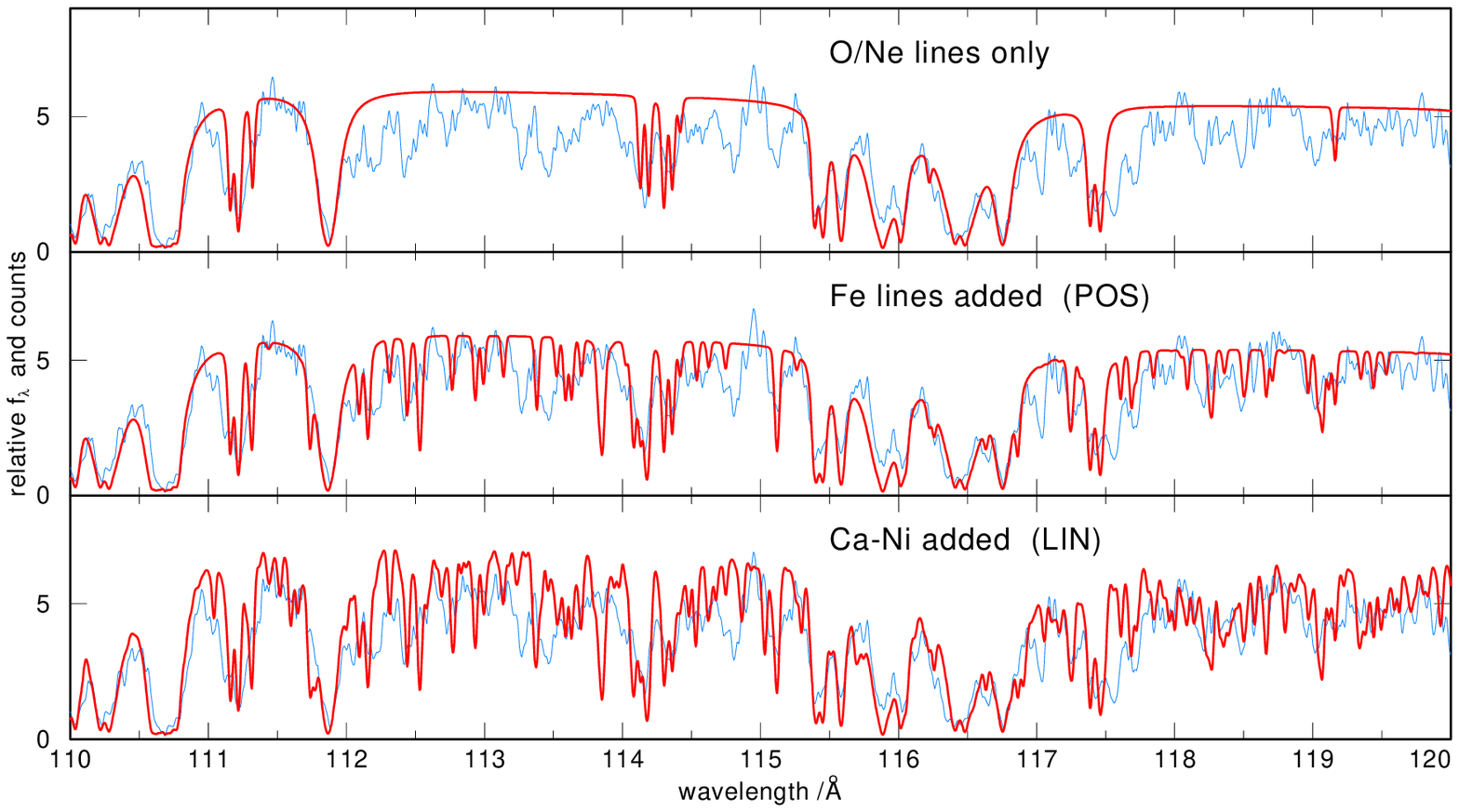}}
  \caption[]{
Effect of line blanketing on the X-ray spectrum of \hh. Top panel: only O and Ne lines
            included in the model. Middle panel: iron lines added. Bottom panel:
            all lines from elements Ca to Ni added. In the bottom panel all lines
            with calculated energy levels, i.e., uncertain wavelength positions,
            are included. This gives an impression of the strong line blanketing.
            It makes individual line identifications of heavy metal lines
            virtually impossible (\Teff=200\,000\,K). Observed count spectrum
            and model spectra (relative flux) are convolved with Gaussians
            with FWHM=0.02\AA\ and 0.03\AA, respectively.
            }
  \label{chandra_fe_group_a}
\end{figure*}

\paragraph{Oxygen} First of all, the observed spectrum does not exhibit \ion{O}{v}
lines. Two relatively unblended \ion{O}{v} lines (at 119.1\AA\ and 124.6\AA) appear very
strong in the 175\,000\,K model and decrease in strength with increasing
\Teff. Even at 200\,000\,K the lines are still quite strong. On the other hand,
at 250\,000\,K all strong \ion{O}{vi} lines generally become too weak. This
clearly suggests \Teff$\approx$200\,000\,K). As we will see later, strong line
blanketing by Fe group elements can severely hamper the detection of individual weak
lines. Fig.\,\ref{chandra_fe_group_a} shows for example how the \ion{O}{v}~119.1\AA\ line
becomes difficult to detect in the respective model spectra.

\paragraph{Neon} We detect several \ion{Ne}{vi} lines, the strongest are located at
122.5,122.7\AA\ and 114.1--114.4\AA. They come out too strong in the
175\,000\,K model and too weak at 250\,000\,K, they fit best at
\Teff=200\,000\,K. \Teff=250\,000\,K can also be excluded because all
\ion{Ne}{vii} lines become too weak.

\paragraph{Magnesium} Increasing \Teff\ in our models from 175\,000\,K to 250\,000\,K
has the following consequences for the line strengths: Lines from \ion{Mg}{v} and
\ion{Mg}{vi} become weaker, \ion{Mg}{vii} strengths are hardly affected (they
are maximal at 200\,000\,K), and \ion{Mg}{viii} lines become very strong. We
point out some individual lines, which altogether suggest \Teff=200\,000\,K:
\ion{Mg}{v}~137.4\AA\ is barely detectable. It would have completely disappeared
if \Teff=250\,000\,K. A broad trough of \ion{Mg}{vii} lines is located near
83--84\AA. It is very nicely fitted by our models, however, it is quite insensitive
against \Teff. More interesting is \ion{Mg}{viii}~74.3--74.4\AA. It is best
fitted by the 200\,000\,K model, while too weak or too strong at 175\,000\,K and
250\,000\,K, respectively.  It seems as if two lines from \ion{Mg}{vi} are
predicted too strong even at 200\,000\,K, namely the multiplets near 123.6\AA\ and
111.7\AA\ but, again, Fe group line blanketing could obscure these features.

As to the magnesium abundance, we calculated a 200\,000\,K model
with the Mg reduced by a factor of 10. The vast majority of the lines
become significantly too weak. In contrast the assumed abundance of 2\% generally gives a
good fit. We estimate the Mg abundance error to 0.5\,dex. In view of the
uncertainty to define the flux continuum we have to accept that
a more accurate determination is not possible.

\begin{figure}[tbp]
  \resizebox{\hsize}{!}{\includegraphics{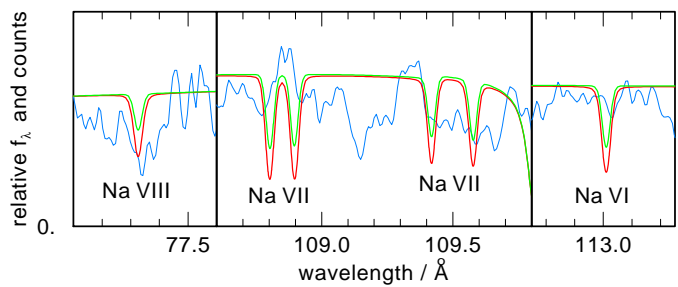}}
  \caption[]{
Synthetic sodium line profiles of three different ionization stages (one \ion{Na}{viii}
            and one \ion{Na}{vi} singlet and two \ion{Na}{vii} doublets) compared to
            the Chandra spectrum. There is no convincing evidence for Na
            lines in \hh\ (\Teff=200\,000\,K, Na abundance 0.3\% and
            0.03\%). Observed and model spectra are convolved with Gaussians
            with FWHM=0.02\AA\ and 0.03\AA, respectively. Tick marks are spaced by 0.1\AA.
            }
  \label{chandra_na}
\end{figure}

\begin{figure}[tbp]
  \resizebox{\hsize}{!}{\includegraphics{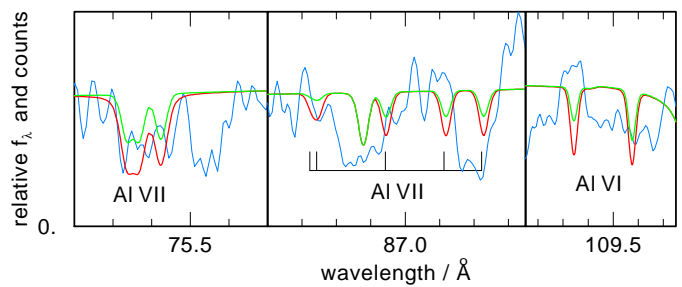}}
  \caption[]{
Synthetic aluminum line profiles of two \ion{Al}{vii} quartets (left two
            panels). The right panel shows two \ion{Al}{vi} lines, a singlet (left
            line) and one
            triplet component. There is no convincing evidence for Al
            lines in \hh\ (\Teff=200\,000\,K, Al abundance 0.1\% and 0.01\%). 
            Observed and model spectra are convolved with Gaussians
            with FWHM=0.02\AA\ and 0.03\AA, respectively. Tick marks are spaced by 0.1\AA.
            }
  \label{chandra_al}
\end{figure}

\subsection{Search for aluminum and sodium}

As already mentioned, the detection of weak individual lines is very difficult
due to strong line blending from Fe group elements. Even careful modeling of
this dense line curtain does not help, because the positions of most Fe group
lines are inaccurate. So we are very conservative in our search for further
species in the Chandra spectrum.

Sodium and aluminum are important elements for the discussion of the evolutionary status of
\hh. Fig.\,\ref{chandra_na} compares calculated sodium line profiles with the Chandra
spectrum. We have selected some of the strongest lines of three ionization
stages, \ion{Na}{vi--viii}. The profiles are computed for two different Na
abundances, 0.3\% and 0.03\%. There is no clear evidence for sodium lines in the
Chandra spectrum. We estimate an upper limit of Na=0.1\%, which is about
30 times solar.

\ion{Al}{vi} and \ion{Al}{vii} show the strongest aluminum lines in the model with
\Teff=200\,000\,K. \ion{Al}{v} and \ion{Al}{viii} are predicted to be much
weaker. Some of these strongest lines are shown in Fig.\,\ref{chandra_al}. Two
models were calculated with abundances of 0.1\% and 0.01\%. Although there seems to be a
good fit by the calculated quartet at 75.2\AA\ (left panel), this is regarded as a chance coincidence with
some other unidentified feature, because other Al lines cannot be identified. We estimate an upper limit of Al=0.1\%, which is about
20 times solar.

\subsection{Search for iron and nickel}

\begin{figure*}[tbp]
  \resizebox{\hsize}{!}{\includegraphics{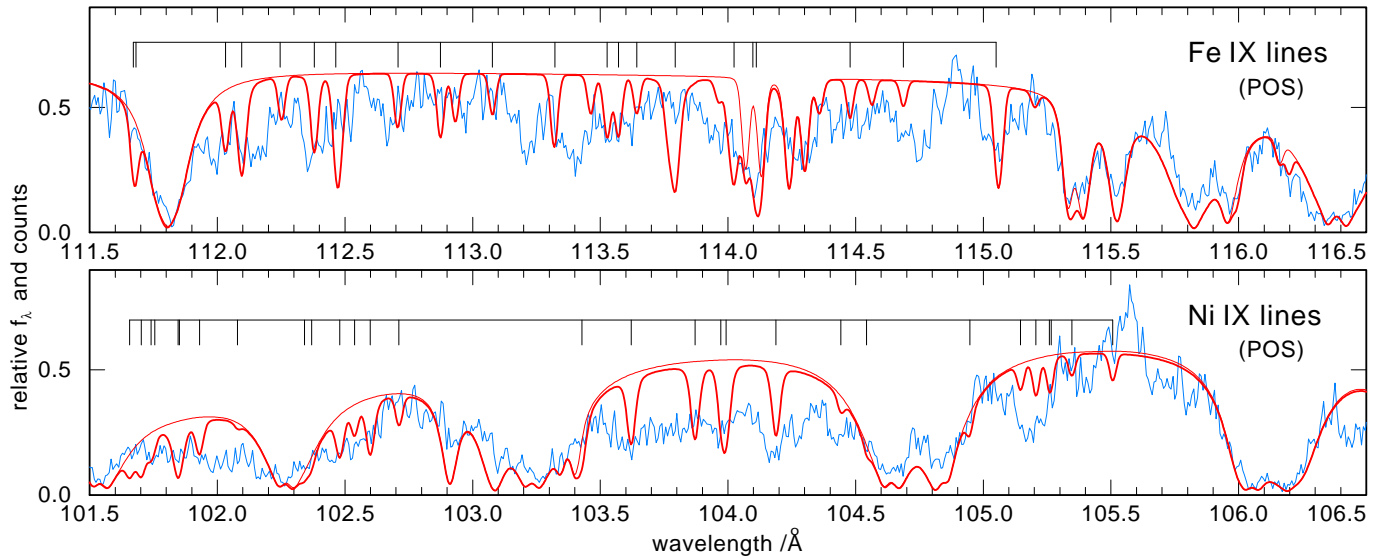}}
  \caption[]{
Synthetic spectra (\Teff=200\,000\,K) with \ion{Fe}{ix} lines (top panel) and
            \ion{Ni}{ix} lines (bottom panel) compared to the
            Chandra spectrum. The POS line lists are used here to enable line
            identifications. Identifying individual lines is impossible, probably
            due to
            strong line blending by other heavy metal lines, as demonstrated in
            Fig.\,\ref{chandra_fe_group_a}. Observed count spectrum
            and model spectra (relative flux) are convolved with Gaussians
            with FWHM=0.02\AA\ and 0.03\AA, respectively.
            }
  \label{chandra_fe_group_b}
\end{figure*}

Figure~\ref{chandra_fe_group_a} shows the strong influence of heavy metal line
blanketing. It becomes clear that the identification of individual lines from
the POS line lists is affected by the many more lines from the LIN lists, which
have uncertain wavelengths. Indeed, it is not possible to identify the
majority of the strong \ion{Fe}{ix} and \ion{Ni}{ix} lines appearing in the
synthetic spectra shown in Fig.\,\ref{chandra_fe_group_b}, hence, we cannot
derive abundances for individual Fe group elements. For all models
including Ca--Ni we
have assumed solar abundances for these elements. This appears to be a good
estimate, because in many parts of the Chandra
spectrum the line blanketed models seem to give a better overall fit to the
observation. So we conclude that the Fe group elements have a total abundance
which is in accordance with solar values, albeit with a large uncertainty.

\subsection{Unidentified absorption features}

Many absorption features in the Chandra spectrum cannot be identified. 
Considering the large number of potential iron group lines
whose exact wavelength position is unknown, this is not surprising. However, let
us comment on about a dozen of the strongest of these features. Some of them can
be attributed to magnesium and neon lines which are not included in our model atoms,
because their high upper level energies would require excessively large model
atoms. These are lines from \ion{Mg}{v} (114.8/115.1\AA), \ion{Mg}{vii} (79.25\AA, 84.64\AA, 87.13/.17\AA,
91.49\AA), \ion{Mg}{viii} (80.89\AA, 84.92\AA), and \ion{Ne}{vi}
(109.03/.07\AA). The strong absorption at 125.1\AA\ remains
unidentified. It cannot be a quartet transition of \ion{Mg}{vi} located
there. It is included in our models but it is very weak. The same applies to a
\ion{Ne}{vi} doublet at this position. A line observed at 86.4\AA\
could be from \ion{Al}{viii}~86.43\AA, but this is a triplet-quintet intercombination
transition and therefore probably too weak to explain the observation. Other strong
unidentified features are located at 93.5\AA\ (a \ion{Fe}{viii} line here is too
weak) and 113.4\AA. We
recall that they could also stem from resonances in bound-free cross-sections.

It is our general impression, supported for example by
 Fig.\,\ref{chandra_fe_group_a}
 that, besides strong O, Ne, and Mg lines, the blanketing by many relatively weak iron
group lines characterizes the Chandra spectrum. Our models show similar
characteristics, except for wavelengths below, say, 100\AA. We attribute this to those
high ionization stages, particularly of Fe and Ni, for which no line lists are
available. We are probably neglecting a large number of \ion{Fe}{xi} and
\ion{Ni}{x-xi} lines at $\lambda<100$\AA, hence, atomic data for these species are badly needed. In
the case of \ion{Ni}{x} we computed the positions of a dozen of resonance lines from Bashkin
\& Stoner (1975) energy levels. They are located around 85\AA\ and one of these lines could
in fact explain the unidentified 86.4\AA\ feature mentioned above.

\section{Summary of spectral analysis}

Let us summarize the properties of \hh:
\begin{eqnarray*}
\Teff &=& 200\,000\,{\rm K} \pm 20\,000\,{\rm K}\\
\logg &=& 8.0 \pm 0.5 {\rm \ \ [cgs]}
\end{eqnarray*}
Element abundances in \% mass fraction:
\begin{eqnarray*}
{\rm C}&=&48 \\
{\rm O}&=&48 \\
{\rm Ne}&=&2 \\
{\rm Mg}&=&2 \\
{\rm Fe-group}&=&0.14\ \ {\rm (solar)} \\
{\rm He}&<&1 \\
{\rm Na}&<&0.1 \\
{\rm Al}&<&0.1
\end{eqnarray*}
The value of \logg\ and abundances for C, O, and Ne were taken from previous work ({\small W91,
WW99}). Estimated errors for abundances are: $\pm$20\% for mass fraction of C and O, and a
factor of 3 for Ne, Mg, and Fe-group. 

Stellar mass and luminosity can be derived by comparing the position of \hh\ in
the $g$-\Teff\ diagram with theoretical evolutionary tracks. We use the post-AGB
tracks of Bl\"ocker (1995) and derive: 
\begin{eqnarray*}
M/{\rm M}_\odot&=&0.836^{+0.13}_{-0.10}\\
\log L/{\rm L}_\odot&=&2.45^{+0.6}_{-0.4}\\
d/{\rm kpc}&=&0.67^{+0.3}_{-0.53}
\end{eqnarray*}
Note that the mass of \hh\ is considerably higher than the mean mass of the PG1159
stars (0.6\,M$_\odot$). The spectroscopic distance was obtained by comparing the measured visual flux
(V=16.24, Nousek \etal 1986) with the flux of the final model
(\Teff=200\,000\,K, \logg\,=\,8): H$_\nu$[5400\AA]=$3.42 \cdot 10^{-3}$
erg/cm$^2$/s/Hz). Interstellar reddening was neglected for this determination,
because it is very low. In fact, the best model fit to the continuum shape
of the UV/FUV spectrum taken with HUT provided E(B-V)=0 (Kruk \& Werner 1998).

\section{Discussion}

PG1159 stars are extremely hot post-AGB stars, many
of them have already entered the hot end of the WD cooling sequence. Generally their
atmospheres are H-deficient and primarily composed of He, C, O, and Ne (typical
abundances are 33\%, 50\%, 15\%, 2\%; Werner 2001). It is now generally accepted that they expose intershell
matter, i.e., material that was located between the H- and He-burning shells in
the former AGB precursor star. In contrast to ``normal'' H-rich post-AGB stars the
PG1159 stars have suffered a late He-shell flash,
during the post-AGB or WD stage. The flash induces
mixing of the intershell (which has a mass of about 0.01M$_\odot$) with the much less massive H-rich envelope (about
$10^{-4}$M$_\odot$). Quantitatively, the intershell element abundances (He, C,
O, Ne) in stellar evolutionary models
are now in good agreement with observations. The observed high O abundances
require strong convective overshoot in order to dredge up oxygen from
the stellar C/O core into the convective intershell (Herwig \etal 1999). 

The high amount of $^{22}$Ne in the intershell is generated by $\alpha$-captures on
$^{14}$N from CNO cycling (see, e.g., Iben 1995, or Lattanzio 2003). Provided the temperature is high enough (which is the
case in relatively massive AGB stars) another $\alpha$-capture on $^{22}$Ne yields $^{25}$Mg or $^{26}$Mg by 
($\alpha$,n) or ($\alpha$,$\gamma$) reactions, respectively. This can explain
the high amount of Mg that we found in \hh, which is the most massive PG1159 star. We also would expect 
to find a Mg enrichment in other PG1159 stars, but the lack of appropriate
spectral lines in the optical/UV prevents its detection.

Besides the production of Ne and Mg in the intershell of AGB stars during the
usual He-shell flashes (thermal pulses), one expects processing of these
elements to Na and Al if they are dredged up into the H-burning shell, namely
through the Ne-Na and Mg-Al cycles: $^{22}$Ne(p,$\gamma$)$^{23}$Na and
$^{25,26}$Mg(p,$\gamma$)$^{26,27}$Al. The high amounts of Ne and Mg in \hh\ on
the one hand and
the fact that we could not detect strong overabundances of Al and Na on the other
hand suggests that these cycles were unimportant for the creation of today's
photospheric chemistry of \hh.

It is unfortunate that we are not able to determine quantitatively the
abundances of individual Fe peak elements. We expect that we should see effects
of s-processing on these elements, shifting away their relative abundances from
the solar values. In fact, it has been found that PG1159 stars are Fe-deficient,
and we suspect this is indeed a result of n-captures on Fe seeds (Miksa \etal
2002, Werner \etal 2003b, Herwig \etal 2003). $^{13}$C is the main neutron
source and it is created by $^{12}$C from He-burning and hydrogen from
the surface layer:
$^{12}$C(p,$\gamma$)$^{13}$N($\beta^+\nu$)$^{13}$C($\alpha$,n)$^{16}$O. Another
source can be the above mentioned $^{22}$Ne($\alpha$,n)$^{25}$Mg reaction.

But how can we explain the He-deficiency of \hh?
Is it just an extreme PG1159 star in which He was completely burnt up to C and
O during the late He shell flash? This can only be a speculation. There are no
appropriate model calculations supporting such a scenario. Another idea
which we already discussed before ({\small W91, WW99}) is that \hh\ has an
evolutionary history completely different from the other PG1159 stars. \hh\ could have
gone through carbon burning, i.e., it would now have a
O-Ne-Mg core and we could see the C/O-rich envelope on top of it. Hence, \hh\ might have been one of the ``heavy-weight''
intermediate-mass stars (8\,M$_{\sun} \lappr M \lappr$ 10\,M$_{\sun}$) which
form white dwarfs with electron-degenerate O-Ne-Mg cores. Evolutionary models by
Iben \etal (1997) predict high Ne and Mg abundances for the C/O envelope, also in agreement with our
analysis. So it is not possible to decide if we are looking onto a C/O core or a C/O
envelope unless, in the latter case, we should see direct
evidence for C-burning. One such piece of evidence would be a strong sodium
($^{23}$Na) enrichment. The $^{23}$Na abundance at the bottom of the C/O
envelope is comparable to that of neon (main isotope $^{20}$Ne) and magnesium
($^{24,25,26}$Mg, see Fig.\,34 of Iben \etal 1997). This is not observed in \hh,
however, we stress the difficulty of giving a tight upper abundance limit for Na
from the present observational data.

The possibility that \hh\ is a O-Ne-Mg white dwarf is remarkable, because
evidence for the existence of such objects is rather scarce (Weidemann
2003). Evidence from single massive WDs is weak, and the most convincing cases are
WDs in binary systems. Strong neon overabundances are found in novae (Livio \&
Truran 1994) or in eroded WD cores in low-mass X-ray binaries (e.g.\ Juett \etal 2001,
Werner \etal 2004).

It is also interesting to note that the supersoft X-ray
source \object{RX\,J0439.8-6809} was suggested to be similar to \hh. This very faint
object (V$\approx$21) has a steep blue and almost featureless optical/UV spectrum, except
for each two \ion{O}{vi} and \ion{N}{v} emission lines in optical spectra (van Teeseling \etal
1999, Reinsch \etal 2002). The presence of nitrogen in completely 3$\alpha$-processed matter
is hard to understand and, instead, it might be possible that the respective optical lines stem
from \ion{C}{v} and \ion{O}{viii}. Such highly ionized emission lines have been
detected in the hottest helium-rich DO white dwarf \object{KPD\,0005+5106} and some PG1159 stars
(Werner \etal 1994) and were attributed to non-photospheric shock-heated emission regions.

\section{Summary}

We have analyzed new FUV and soft X-ray spectra of the unique object \hh. We
confirm its exotic chemical composition, which is dominated by C and O. We
confirm the high Ne abundance and find a similarly high abundance of Mg. This
chemistry either reflects that of the core of a C/O white dwarf or the C/O
envelope of a white dwarf with a O-Ne-Mg core. It therefore remains unclear if
\hh\ has gone beyond 3$\alpha$ burning through a subsequent C burning phase or not. 
In any case, the origin of the He-deficiency remains obscure. \hh\ could be
an extreme PG1159 star which -- in contrast to the other stars of this group -- has
for some unknown reason burned up its helium completely. Alternatively,
\hh\ could have burned carbon, now being a
O-Ne-Mg white dwarf. Some unidentified mechanism (C shell flashes?) may be responsible for the loss
of helium by ingestion and burning in deep hot layers. Interestingly, Iben and
collaborators have predicted that such C-burning stars in the super-AGB phase could 
loose their H- and He-rich envelopes by a radiation driven superwind (Ritossa \etal
1996). \hh\ might resemble the result of such a scenario.

A surprisingly rich photospheric absorption line spectrum in the soft X-ray regime has been
revealed by our Chandra observation. Although the overall flux distribution
cannot be explained by a single model with a particular temperature, the
ionization equilibria of O, Ne, and Mg suggest that \Teff\ is slightly higher than
determined in previous analyses (200\,000\,K$\pm$20\,000\,K). This makes \hh\ the
hottest known post-AGB star and white dwarf ever analyzed in detail with model
atmosphere techniques.

\begin{acknowledgements}
FUSE and Chandra data analysis in T\"ubingen is supported by the DFG under grant We\,1312/30-1.
TR is supported by the DLR under grant
50\,OR\,0201. MAB is supported by the Particle Physics and Astronomy Research
Council, UK. JWK is supported by the FUSE project, funded by NASA contract
NAS5-32985. We thank Dr.\
Kramida (NIST)
and Prof.\ Kunze (University of Bochum) for their advice on the
\ion{Ne}{vii}~973\AA\ line identification. This research has made use of the SIMBAD
Astronomical Database, operated at CDS, Strasbourg, France.
\end{acknowledgements}


\begin{thebibliography}{}

\bibitem{BS75} Bashkin, S., \& Stoner, J.O. Jr. 1975, Atomic Energy Levels \&
        Grotrian Diagrams, Vol.\ 1, Amsterdam, North Holland

\bibitem{Blo95} Bl\"ocker, T. 1995, A\&A, 299, 755

\bibitem{Her99} Herwig, F., Bl\"ocker, T., Langer, N., \& Driebe, T. 1999, A\&A, 349, L5

\bibitem{Her03} Herwig, F., Lugaro, M., \& Werner, K. 2003, in Planetary Nebulae: Their Evolution and Role in the Universe,
        eds.\ S.\,Kwok, M.\,Dopita, R.\,Sutherland, IAU Symp.\ 209, ASP, p.\,85 

\bibitem{Iben95} Iben, I. Jr. 1995, Phys. Rep., 250, 2

\bibitem{Iben97} Iben, I. Jr., Ritossa, C., \& Garcia-Berro, E. 1997, ApJ, 489, 772

\bibitem{John71} Johnston, W.D., \& Kunze H.-J. 1971, Phys. Rev. A, 4, 962

\bibitem{Jue01} Juett, A.M., Psaltis, D., \& Chakrabarty, D. 2001, ApJ, 560, L59

\bibitem{KrW98} Kruk, J.W., \& Werner, K. 1998, ApJ, 502, 858

\bibitem{Ku91} Kurucz, R.L. 1991, in Stellar Atmospheres: Beyond
	Classical Models, eds. L.\,Crivellari, 
	I.\,Hubeny, D.G.\,Hummer, Kluwer, Dordrecht, NATO ASI Series C, 341, 441

\bibitem{Lang83} Lang, J. 1983, J. Phys. B, 16, 3907

\bibitem{Lat03} Lattanzio, J. 2003, in Planetary Nebulae: Their Evolution and Role in the Universe,
        eds.\ S.\,Kwok, M.\,Dopita, R.\,Sutherland, IAU Symp.\ 209, ASP, p.\,73

\bibitem{Lin72} Lindeberg, S. 1972, Uppsala Univ.\ Inst.\ Phys., Report UUIP-759, 1

\bibitem{Liv94} Livio, M., \& Truran, J.W. 1994, ApJ, 425, 797

\bibitem{Mik02} Miksa, S., Deetjen, J.L., Dreizler, S., Kruk, J., Rauch, T.,
	\& Werner, K. 2002, A\&A, 389, 953

\bibitem{Moo00} Moos, H.W., Cash, W.C., Cowie, L.L., \etal 2000, ApJL, 538, 1

\bibitem{Nousek96} Nousek, J.A., Shipman, H.L., Holberg, J.B., \etal 1986, ApJ, 309, 230

\bibitem{Nug83} Nugent, J.J., Jensen, K.A., Nousek, J.A., \etal 1983, ApJS, 51,1 

\bibitem{Rau03} Rauch, T., \& Deetjen, J.L. 2003, in
        Stellar Atmosphere Modeling, eds.\ I.\,Hubeny, D.\,Mihalas,
        K.\,Werner, ASP Conf. Ser., 288, 31

\bibitem{Rei02} Reinsch, K., Beuermann, K., \& G\"ansicke, B.T. 2002, in The Physics of
  Cataclysmic Variables and Related Objects, eds. B.T.\,G\"ansicke, K.\,Beuermann,
  K.\,Reinsch, ASP Conf. Ser., 261, 653

\bibitem{Rit96} Ritossa, C., Garcia-Berro, E., \& Iben, I. Jr. 1996, ApJ, 460, 489

%\bibitem{Rumph94} Rumph, T., Bowyer, S., \& Vennes, S. 1994, AJ, 107, 2108

\bibitem{Sah00} Sahnow, D.J., Moos, H.W., Ake, T.B., \etal 2000, ApJL, 538, 7

\bibitem{Sea94} Seaton, M.J., Yan, Y., Mihalas, D., \& Pradhan, A.K. 1994, MNRAS, 266, 805

\bibitem{van99} van Teeseling, A., G\"ansicke, B.T., Beuermann, K., Dreizler,
  S., Rauch, T.,\&  Reinsch, K. 1999, A\&A, 351, L27

\bibitem{Weid03} Weidemann, V. 2003, in White Dwarfs, eds.\ D.\,de
	Martino, R.\,Silvotti, J.-E.\,Solheim, R.\,Kalytis, NATO Sci. Ser. II, Kluwer, Vol.\ 105, 3

\bibitem{KW91} Werner, K. 1991, A\&A, 251, 147 ({\small W91})

\bibitem{KW96} Werner, K. 1996, ApJ, 457, L39

\bibitem{KW01} Werner, K. 2001, in Low Mass Wolf-Rayet Stars:
	Origin and Evolution, eds. T.\,Bl\"ocker, L.B.F.M.\,Waters,
	A.A.\,Zijlstra, Ap\&SS, 275, 27

\bibitem{WW99} Werner, K., \& Wolff, B. 1999, A\&A, 347, L9 ({\small WW99}) 

\bibitem{WHH91} Werner, K., Heber, U., \& Hunger, K. 1991, A\&A, 244, 437

\bibitem{WHF94} Werner, K., Heber, U., \& Fleming, T. 1994, A\&A, 284, 907

\bibitem{Wetal03a} Werner, K., Deetjen, J.L., Dreizler, S., Nagel, T., Rauch,
	T., \& Schuh,
	S.L. 2003a, in
        Stellar Atmosphere Modeling, eds.\ I.\,Hubeny, D.\,Mihalas,
        K.\,Werner, ASP Conf. Ser., 288, 31

\bibitem{Wetal03b} Werner, K., Deetjen, J.L., Dreizler, S., Rauch, T., \& Kruk, J.W. 2003b,
	 in Planetary Nebulae: Their Evolution and Role in the Universe,
        eds.\ S.\,Kwok, M.\,Dopita, R.\,Sutherland, IAU Symp.\ 209, ASP, p.\,169
 
\bibitem{Wetal04} Werner, K., Nagel, T., Dreizler, S., \& Rauch, T. 2004, 
	in Compact Binaries in the Galaxy and Beyond, IAU Coll.\ 194, in press
	(astro-ph/0312561)

\bibitem{Yo03} Young, P.R., Del Zanna, G., Landi, E., Dere, K.P., Mason, H.E.,
        \& Landini M. 2003, ApJS, 144, 135

\end{thebibliography}
\end{document}